\documentclass[oneside,masters,etd]{WSUclass}

\usepackage{import}

\usepackage{preamble}

\usepackage[utf8]{inputenc} % Remove warning on ascii conversion
\usepackage[T1]{fontenc} % Remove warning on ascii conversion
\usepackage[refsection=part,citestyle=ieee,style=numeric,natbib=true,backend=biber, sorting=none]{biblatex}
\usepackage{hyperref}

\usepackage{fmtcount}
\makeatletter
\renewcommand{\@makechapterhead}[1]{\vspace *{40\p@ }{\parindent \z@ 
\raggedright \normalfont \ifnum \c@secnumdepth >\m@ne \Huge \bfseries 
\@chapapp \space \Numberstring{chapter} \vskip 10\p@ \fi #1\par \nobreak \vskip 30\p@ }}
\makeatother

\begin{document}

\hypersetup{breaklinks=true}

 % Start page counting in roman numerals
 \frontmatter

 % This command makes the formal preliminary pages.
 % You can comment it out during the drafting process if you want to save paper.
 \makepreliminarypages
 \doublespace
 % Make the table of contents.
 \tableofcontents
 \thispagestyle{plain}

 % Make the list of tables
 \mylistoftables
 \thispagestyle{plain}
 
 % Make the list of figures
 \mylistoffigures
 \thispagestyle{plain}

  % This page is OPTIONAL. To remove, comment out and \dedicationpage in diss.tex
 \dedicationpage
 \clearemptydoublepage

 % Start regular page counting at page 1
 \mainmatter

% OK. Everything is set up. Type your thesis here.
\addchapheadtotoc
\chapter{Introduction}
\section{Review}
Medical image processing is one of the areas of research that physicians and scientists are paying particular attention to today. Therefore, digital imaging studies are performed to gain access to detecting and displaying organs of the body, especially the affected organs. Advances in medical image analysis have greatly improved the treatment of diseases. Nowadays, physicians can look inside the body for treatment and follow the process of changes in body organs and employ different mechanisms to follow the treatment process more efficiently than ever before. With automated, efficient, sustainable, inclusive, and responsive tools for real-time systems, they can have a support system to validate their views and improve treatment by conducting multiple surveys to measure various parameters. To empower the medical system with this sort of capabilities, we need knowledge in several areas, including Image processing and machine vision, machine learning, pattern recognition, and expert systems \cite{osareh2004automated}.
Since diseases such as Glaucoma, as well as symptoms of many diseases, such as diabetics and hypertension, appear in the retina, examining and processing retina images is one of the most valuable tasks in medical image processing. Detection of vein thrombosis disease, measurement of the degree of curvature of the vessels in hypertension, and identification of the start of Glaucoma, based on the characteristics of retinal blood vessels, indicate the great importance of the analysis of this layer of tissue of the eye. Moreover, vessel segmentation is a pre-processing task to identify other areas of the retina and bright and dark abnormalities \cite{lam2010general}.
Arteries and veins have features such as width, colour, curvature, and blur. Also, the pixels of the black and red contact parts and the pattern of the small arteries are useful in identifying diseases. Besides, they have many applications in the treatment process and clinical trials. Consequently, by detecting the retinal vascular structure accurately, such information can be obtained from the retina images\cite{hoover2000locating}.
\section{Challenges and Incentives}
Non-automatic segmentation of the retinal blood vessels is very time-consuming, tedious, and requires specialised training and skills. Automatic detection of retinal blood vessels can be used as an initial step in retinal disease identification algorithms to build and deploy a computerised system for the treatment of eye injuries. 
With the development of eye image processing systems, a large number of pathological images can be collected for analysis. Processing these images in computerised medical systems requires rapid and sufficiently efficient segmentation algorithms to process images obtained from different imaging systems and imaging conditions \cite{fraz2012blood}.
Retinal vessel segmentation is based on the properties of the vessels and other parts of the retina itself. The alignment and grey level of the vessels never change abruptly; they are locally linear, and their intensity changes slowly. The vessels are interconnected and form a tree structure in the retina. The bright and dark anomalies, the pixels of arterial segments, and the similarity of some background regions to the arteries make their detection complex. Like most medical imaging problems, low contrast images are another challenge in the process of segmentation. Also, retinal vessels sometimes exhibit a large amount of light reflection, which occurs more frequently in arteries \cite{fraz2012blood}.
In this dissertation, we sought a method to extract retinal vessels in an efficient, inclusive, and responsive manner. In addition, since retinal images produced in medical centres usually contain a variety of abnormalities, and these abnormalities are manifested in response to vessel extraction methods as False-Positive pixels, in this study, two different methods for improving the performance of automatic vessel detection subsystem is suggested to eliminate such abnormalities before the vessel detection process begins.
\section{A Brief Overview of the Methods}
Figure 1.1 shows an overview of the two methods of this thesis for the detection of retinal vessels. In general, the input retinal image is processed automatically, and in the end, a binary image, which contains the structure of the vein, is obtained.
At the beginning of both methods, two pre-processing functions are performed on the green channel of the retina image. First, the boundaries of the field of view (FOV) are expanded to reduce the likely error that occurs in the image boundaries during the vessel segmentation algorithm. Then, to enhance the contrast, the image is taken to another space, called the Weber space.

\begin{figure}[ht!]
\begin{center}
\includegraphics[scale=0.5]{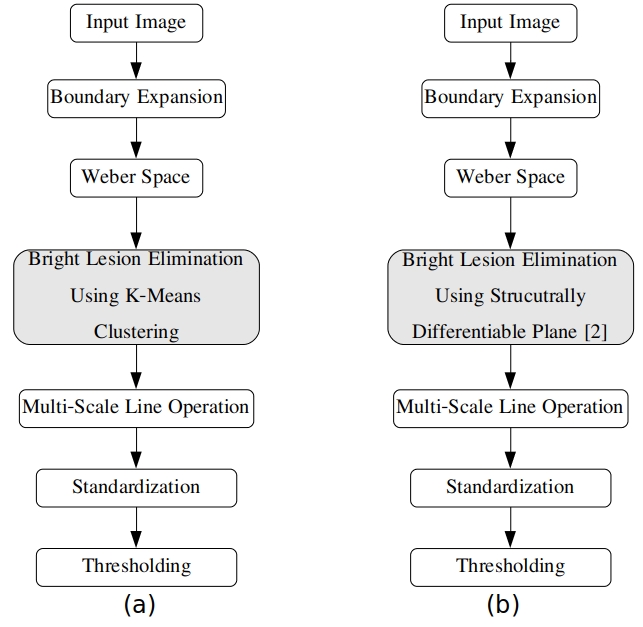}
\end{center}
\caption{The presented methods in one view. a) Method 1. b) Method 2}
%Source:
\label{image-Methods}
\end{figure}

As shown in the diagram of the Fig. \ref{image-Methods}, the difference between the two methods is only in how they perform in attenuating bright lesions (Exudates) to increase the efficiency of the vessel detection algorithm.

The process of reducing bright lesions in Method 1 is based on a machine-learning algorithm, called K-means Clustering, and in Method 2, with the help of an optimization process and obtaining a structurally differentiable (SD) plane using a Total Variation (TV) filter.

The most important feature of the blood vessels is their linear structure. Thus, by identifying lines in the retina image, the vessels can be identified alongside ignorance of other nonlinear structures, such as dark anomalies (Microaneurysm, Hemorrhage), optic disc, and fovea (macula). Therefore, after reducing the bright lesions, an algorithm is employed to detect the linear structures of the vessels using a multi-scale line operator. Finally, the response of the multi-scale line operator is thresholded.

To evaluate the performance of the methods, images of the two datasets, DRIVE \cite{niemeijer2004comparative} and STARE \cite{goldbaum1975structured} were exploited, and the system performance was obtained based on segmentation accuracy criteria, the area under the ROC (Receiver Operating Characteristic) diagram, and response time.

\begin{figure}[ht!]
\begin{center}
\includegraphics[scale=0.5]{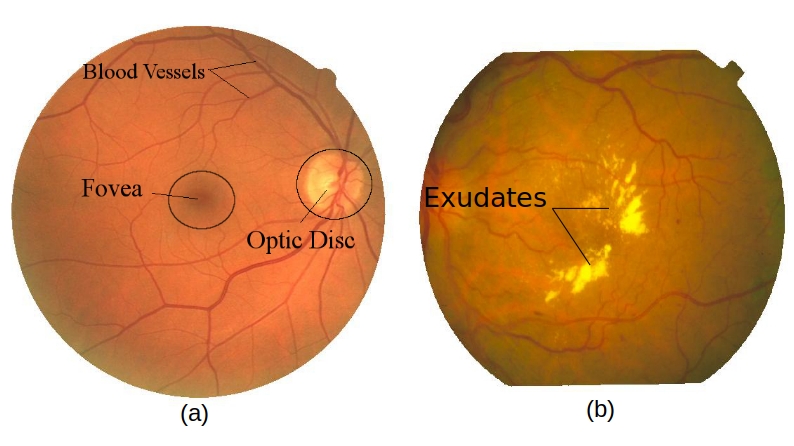}
\end{center}
\caption{Retina Image. a) Retina image and its main parts (image number 16 of DRIVE dataset). b) A sample of retina having anomaly in Fovea-Macula area (image number 1 from STARE dataset).}
%Source:
\label{image-retina-parts}
\end{figure}

\section{Retina Image}
Retinal Fundus Imaging, compared to other imaging modalities, such as Fluorescein Angiography, provides richer color images, it is less expensive and has no side effects. These images contain the current state of the retina. Therefore, the physician can monitor changes occurring in several time frames, in addition to being able to perform further examinations on the patient's retina. 

Resolution is one of the features to describe color retina images. Some cameras can provide \(2336\times 3504\) images or the like. For practical application with limited processing resources the resolution of these images is reduced to \(700 \times 700\), \(605 \times 480\), and so on, in order to make the processing methods more time-efficient.

Section (a) of Fig.\ref{image-retina-parts} shows a retina. The retina has three main parts of the vascular structure, the optic disc and the Fovea-Macula section. The vascular structure is a combination of veins and arteries on the retina with their branches. The optic disc can be seen in the shape of a light yellow circle where the veins intersect. In the middle of the image is Fovea, which is very sensitive to light and provides a direct and clear vision. In the middle of Fovea is the Macula. The Fovea-Macula section in the retina of the healthy eye is usually a dark circle.

Diseases such as Hpertension, high blood lipids, Diabetic, and Glaucoma can cause disruption to the structure of each of the retinal segments. In part (b) of Figure 1.2, an example of a pathological retina with bright lesions is shown.

\section{Thesis Structure}
In Chapter 2, a background of the previous work in the field of retinal vessel segmentation . In Chapter Three, we will review the retina image pre-processing algorithms and introduce segmentation and detection algorithms. Chapter 4 presents the proposed methods with details. Chapter 5 analyzes the effectiveness of the proposed methods, and Chapter six contains conclusions and suggestions for future work.
\chapter{Literature Review}
Because of the importance of vessel segmentation in retinal images, a lot of research has been done on the automatic segmentation of this structure from the retina with the help of computers. Previous methods can be categorized from different perspectives. In a general point of view, the methods are either supervised or unsupervised. Furthermore, from a more specific perspective, they can be classified based on their main segmentation approaches: pattern recognition, window-based, tracking, morphology, multi-scale, or model-based. In addition, another more practical categorization is whether the overall method takes care of the pathological abnormalities in retina images \cite{fraz2012blood}. Most of the proposed methods work on color retinal images, and angiographic images have been used in a small number of studies \cite{espona2006vascular}.

In the course of this chapter, we review a number of previous works on the identification and segmentation of vessels in retinal images.

\section{Overview of Previous Works}
Unsupervised methods look for patterns and features inherent in vessel structures in retinal images. These patterns and features can be used to identify whether or not a pixel is a vessel. Training or ground truth data, used in the construction of such methods, is more commonly used to know the proper location of vessels and to measure the segmentation performance. 

Matching filters, such as Gaussian filter, second derivative, Gaber and steerable filters, have been able to provide a good pattern of vessel cross section. In \cite{chaudhuri1989detection}, Chadori et al. measured the similarity between an inverse Gaussian pattern and the vessels cross section using a matching filter. Hoover, in \cite{hoover2000locating}, has improved this approach by adding a process to eliminate noise. In \cite{gang2002detection}, after estimating the parameters of the Gaussian pattern of vessels cross section, a second-order modified-amplitude Gaussian filter is proposed to identify the position and direction of the vessels.

In \cite{fielder2007automatic}, the Laplacian of Gaussian (LoG) filter optimized for edge detection, Otsu's thresholding, and skletonization and thinning are enhanced for vessel segmentation.

In \cite{tolias1998fuzzy}, a tracking method based on region growing is used. The seeds for this tracking method has been expanded on the basis of the intensity concavity of vessels.
A statistical-based tracking method is used in \cite{yin2010bayesian}. In this work, a statistical model is optimized to identify the pixels on the edge of the vessels. This model combines the features of the grey and geometric levels of the vessels. The pixels on the vessel border are revealed by the Bayesian method and a probability criterion in an iterative process.

In \cite{hatanaka2004automated}, with the help of a two-tier filter and filter response thresholds, the candidate pixels are identified. Then, the isolated False-Positive pixels and far from the optic disc are removed. After identifying the vessel centre lines, a tracking method is used to detect the vascular network. In this work, the optic disc is removed from the image in a simple way before detecting and tracking the vessels. It is clear that this method does not examine the vessels within the optic disc.

In supervised methods, with the help of training data, an algorithm learns vessel pattern. These training data are images that the ophthalmologists have labelled; of course, there is no complete agreement among physicians on the labelling of the vascular structures \cite{fraz2012blood}. In a supervised method, they categorize data based on several attributes derived from the labelled data. Therefore, one of the prerequisites of these methods is the availability of labelled images, which is hardly possible in practical applications. On the other hand, since these methods are based on preconfigured parameters, they are more efficient than unsupervised methods and provide excellent responses to healthy retina images.

In \cite{staal2004ridge}, one of the earliest methods of vessel segmentation has been performed using a supervised learning approach. To train a classifier, several features are used to separate the pixels into vessel and Non-vessel. The method provides good response on healthy retina images. But many of the features used, such as the Laplacian of Gaussian filter, used to detect the intensity of vessels, require that portions without a vessels of the image be very smooth. If there is a bright lesion in the image, this method detects a large number of False-Positive pixels.

In \cite{marin2010new}, a seven-characteristic vector of grey levels (of neighbourhood properties) and moment invariant features (related to vessel linearity property) are considered. Then, a Neural Network classifier segments the pixels based on their feature vectors.

In \cite{soares2006retinal}, the authors used the feature vector obtained from the Gabor filter. A learning process is used to adjust the parameters of this filter. These parameters are related to low-frequency structures in cross sections of vessels. This technique identifies vessels in retinal images. However, as it looks for vessels based on their low frequency characteristic, it fails to identify the vessel pixels along with the bright lesions.

In \cite{osareh2009automatic}, for each pixel of the image, a feature vector is obtained based on the direction and size parameters of the Gaber filters. With these features, two classifiers have been trained to separate the pixels into vascular and non-vascular categories. Due to the optimization of the applied parameters, this method works well for both pathological and healthy image groups. The same Gabor filter capability has been used in \cite{siddalingaswamy2010automatic} to reduce background noise.

In \cite{santhi2011locating}, fuzzy classifiers have been used to extract vessels. In \cite{xu2010novel}, first the large blood vessels were segmented by an Adaptive Local Thresholding, and then the small vessels were detected by a Support Vector Machine. In the end, by tracking these small veins, the whole vascular structure is estimated.
Threshold probing techniques have also been shown to be effective in the vessel segmentation. In \cite{hoover2000locating}, after obtaining the Matching filter response, the image pixels are grouped into vessel and non-vessel classes based on the neighborhood and region based features, and then by reducing the threshold value.

From the morphological point of view, the most obvious feature of the vascular network is that it is made up of interconnected linear segments \cite{fraz2012blood}. Therefore, in many works, morphological operators have been used to operate on retinal images. In \cite{kumari2010blood}, they used the Wiener filter and the morphological opening and closing operators to extract vessels. The signal-to-noise ratio of the two methods showed that morphology operators outweigh the other method. In \cite{ding2011approach}, a fuzzy clustering  with morphological operators are applied together.

In \cite{ricci2007retinal}, an application of line operators is proposed to construct feature vectors for pixel and then separating them with support vector machines. A line operator passes through every pixel of the green channel image, in several directions, and the new intensity value of that pixel will become the average values of the pixels intensities along the line. The thresholded response of the resulting image has shown the effectiveness of the method in extracting vessels. In this study, two perpendicular line detectors have also been used to extract feature vectors to further be classified by a support vector machine. The same line operator has been used in a multi-scale fashion to detect vessels in \cite{nguyen2013effective}.

The methods investigated so far have performed well on healthy retina images, but most do not respond well to pathological retina images. Because in these methods, usually non-vascular portions in the image are assumed to be smooth, which is incorrect given the presence of bright and dark anomalies. Pathology techniques are approaches in which vessels are segmented, with special attention to retinal abnormalities. A small portion of the available methods have paid special attention to retinal anomalies in the process of vessel extraction.

In \cite{zana1997robust}, an algorithm based on morphology and line detection is proposed for the identification of vessels in noisy angiographic images. Also, a geometrical model is used to extract the vessels from their surrounding regions. In this work, before the vessels are segmented, the bright regions, with the highest grey level, have been isolated with the aim of eliminating microaneurysms.

In \cite{leandro2001blood}, they have also used morphology to find vessels and eliminate microaneurysms, and then they used a Wavelet filter to reduce noise and signify vessels.

In \cite{mendonca2006segmentation}, the \(L \times a \times b\) color space is used to eliminate bright lesions. A method based on the Divergence Vector Field is also introduced in \cite{lam2008novel}, in order to eliminate bright lesions by obtaining degree of smoothness of the vessel-free sections in different directions. Although this method works well for bright lesions, it requires a thresholding process for artifacts. A large threshold eliminates more artifacts near bright lesions but also neglects many vessel pixels.

In \cite{lam2010general}, a method is proposed to eliminate dark and clear abnormalities along with the identification of vessels. In this work, a method is used to model multi-concavities. This method is more efficient than other methods of analyzing pathological images.

In \cite{ying2011modeling}, a framework for modeling the retinal vessels, to obtain its statistical properties. In addition to identifying the vessels, other parts of the retina have also been identified with the help of the vascular  structure.

Some of the previous work has only been done on parts of the vascular structure. In \cite{aibinu2010vascular}, a method called "Combined Cross-Point Number (CCN)" is proposed for identifying pixels in vessel intersections and bifurcations.

Although numerous work has been done to extract vessels, from healthy and pathological images it can still be considered as one of the most difficult computer vision tasks \cite{lam2010general}. In addition, only a small number of these works have been able to perform well in detecting boundaries, shallow vessels, and trade-off between sensitivity and specificity \cite{ying2011modeling}. Also, in many of them, there are crucial parameters that need to be optimized. The optimality of some of these parameters for segmentation in one image set may not be effective for another.

In this chapter, some of the previously proposed methods for detecting retinal blood vessels, are reviewed and their capabilities, shortcomings and challenges presented.

In this thesis, we have attempted to detect vessels according to the pathological aspects of retinal images. Proposed methods can be categorized into unsupervised and pathological categories. Utilizing machine learning processes, geometrical and mathematical properties of vascular structures and anomalies to eliminate bright and dark anomalies and identifying linear vascular structures are features of the proposed methods.

Chapter 3 discusses how retinal colour images are processed to provide the conditions needed to identify and isolate vessels.
\chapter{Theoretical Foundations}
In this chapter, after reviewing the retina image datasets, the image preprocessing and image segmentation tools used in this thesis are presented. The reasons for using each of these tools and how they work will be discussed later in Chapter 4.

\section{Retina Image Datasets}
In this thesis, we used the images of two publicly available retina image datasets DRIVE and STARE to evaluate the efficiency of the vessel segmentation procedures.

The DRIVE database contains two batches of twenty training and test images. Each training image has one and each test image has two corresponding labelled images. The STARE dataset has twenty images. Ten of them are pathological retina images and the other ten are normal retina images. Each image in this set has two corresponding labelled images. Images in DRIVE have fewer anomalies, but the contrast is better for STARE images.

The two methods are applied to the sixty images from the two datasets, and the results are compared with their labelled images. Figure 3.1 shows an image from the DRIVE dataset and its corresponding labelled image.

\begin{figure}[ht!]
\begin{center}
\includegraphics[scale=0.5]{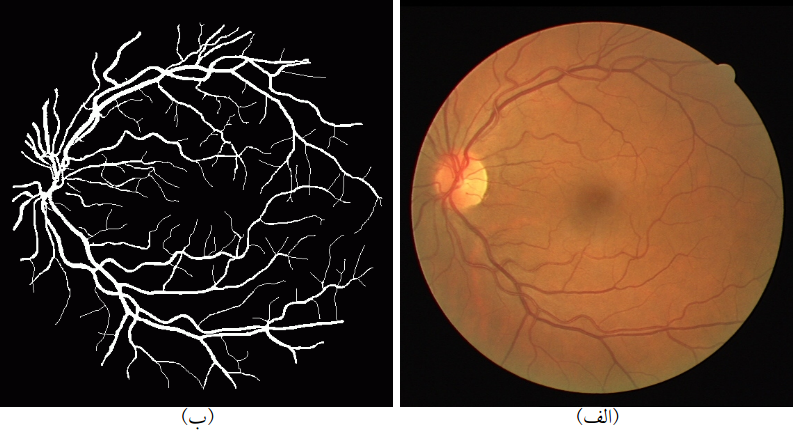}
\end{center}
\caption{A retina Image from DRIVE dataset (a), and its corresponding labelled image (b.)}
%Source:
\label{3.1-image-DRIVE}
\end{figure}

\section{Binary Mask}
The binary mask of a retina image (Part (b) of Figure 3.2) is an image that is the same size as a channel of the original retina image, in which regions consisting of pixels of intensity 1 represent the the foreground of the original image (field of view), and pixels with intensity 0, indicate the background area (the darker portion around the retina in part an of Figure 3.2) [32]. With the precise positioning of the field of view, one can only focus on its pixels and avoid unnecessarily increasing processing time.

\begin{figure}[ht!]
\begin{center}
\includegraphics[scale=0.5]{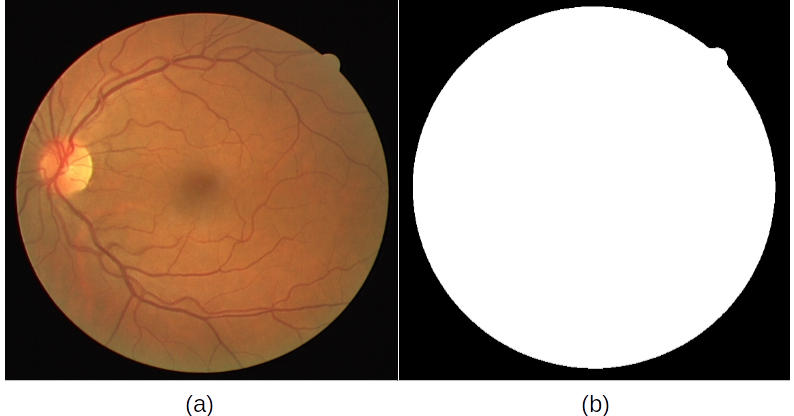}
\end{center}
\caption{A retina Image (a) and its corresponding mask (b.)}
%Source:
\label{3.2-image-retina-mask}
\end{figure}

In this regard, the retina image can be easily separated from the background by converting its RGB image to an HSI image. In this new colour space, there is a layer (channel) only to separate the intensities. Since the background is darker than the FOV, simple thresholding can be used for this purpose. Then, to eliminate any possible noise, an averaging filter is used, and to eliminate the pixels on the edges, the morphological erosion operator with the structuring element \(5 \times 5\) is applied.

\section{Green Channel of Retina Colour Image}
As with many of the previous works reviewed in Chapter 2, in this study, the input image is the green channel of the retina colour image, in which the grey level of each pixel is taken to the interval [0, 255]. The reason for this is the higher resolution of the blood vessels compared to the background in this layer of the image. Figure 3.3 shows all three layers of the retina image.
\begin{figure}[ht!]
\begin{center}
\includegraphics[scale=0.5]{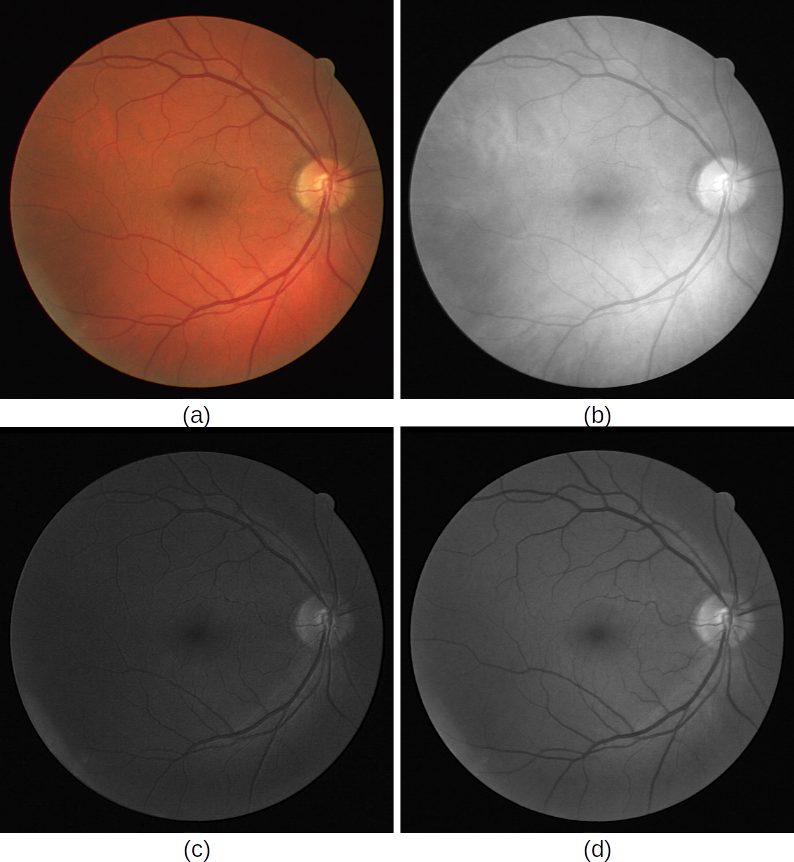}
\end{center}
\caption{An original retina colour image (a), the red channel (b), green channel (c), and the blue channel (d.)}
%Source:
\label{3.3-retina RGB-channels}
\end{figure}

\section{Pre-processing}
Sometimes retina images have blurry light and contrast. Since the processes of the applied segmentation methods are performed in small neighbourhoods, there is no need for a general pre-processing, such as illumination equalization or histogram equalization (although these methods were implemented and the results didn't make a considerable difference to the final accuracy). Also, if we can model the vessel structure well, we can ignore this issue in images. On the other hand, since these optical inhomogeneities do not have the same pattern in all images, applying an algorithm with constant parameters to smooth them all reduces the generality of the methods. Such parameters including the size of the window in the illumination equalization method responds better to one set of images than others. In this thesis, the pre-processing procedures do not require parameters to be adjusted.

\subsection{Transforming to Weber Space}
To improve retinal image contrast, we can use an image space transfer based on Weber's law \cite{majumder2007perception}. Weber calculated the relationship between physical intensity magnitude of an image and the intensity perceived by eye (perceptive intensity). Briefly speaking, Weber's law states that there is a logarithmic relationship between what an image represents and what our eyes receive. In \cite{lam2010general} and \cite{shen2002weber}, how to compute the amount of illumination received using the Weber space, is proposed. In fact how the conversion of the input image to the received image (Weber) is defined as follows:

\begin{equation} \label{weber-formula}
v^0(z) = \frac{ln(1+I^0(z))}{k}    
\end{equation}

Where \(k\) is a constant number and is considered to be equal to 1. If another value is used for k, it will have no effect on the segmentation results as it is neutralized by normalization processes and other parameters. Studies have shown that this transfer actually enhances the efficiency of the optimization process of the Total Variation model (see Section 3.5), thereby enhancing the quality of the processed image and increasing the potential for more accurate vessel segmentation from the background \cite{lam2010general}, \cite{shen2002weber}. In this thesis, we have used this transition as a pre-processing step in both methods of segmentation (Chapter 4). The appendix provides details of the law and the Weber space.

It follows from the formula \ref{weber-formula} that if we give the intensity of image pixels with values in the interval [0, 255], the intensity of the output image pixels will be in the interval [0, 5.55]. Figure 3.4 shows an example of the result of this transform. In this figure, we have scaled the Weber image to the interval [0, 1] to be able to display the result.

\begin{figure}[ht!]
\begin{center}
\includegraphics[scale=0.5]{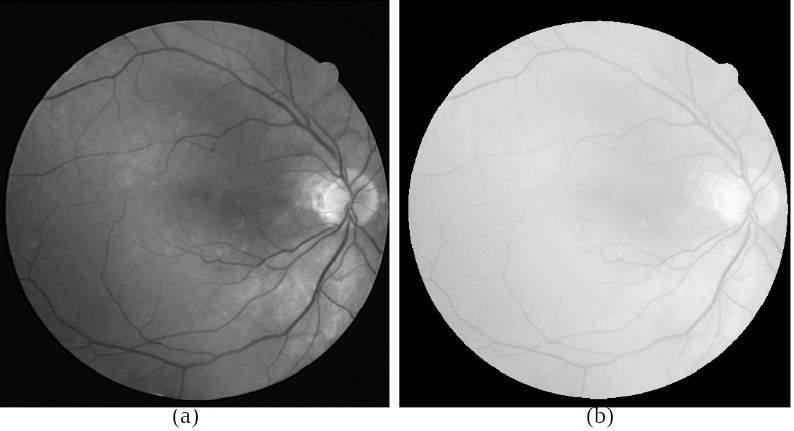}
\end{center}
\caption{Transforming the green channel of retina image to Weber space. (a) The grey image. (b) The image in Weber space (the image is scaled to interval [0,1].}
%Source:
\label{weber-transform-image}
\end{figure}

\subsection{Expanding the Boundaries of the Field of View}
\begin{figure}[ht!]
\begin{center}
\includegraphics[scale=0.5]{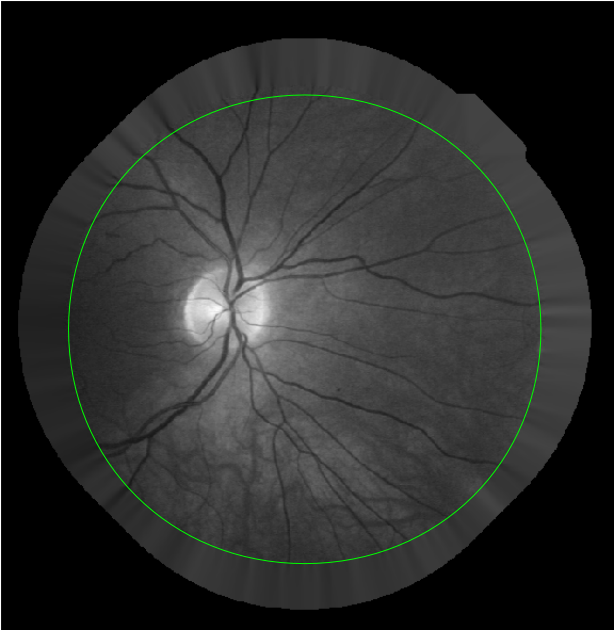}
\end{center}
\caption{An example of boundary expansion (the green line is the edge of the FOV}
%Source:
\label{boundary-expansion-image}
\end{figure}

Another challenge in identifying retinal vessels is the similarity of the boundaries of the field of view to the vessels. These borders, like vessels, are linear and edge-like. If they are not ignored during segmentation, many False-Positive pixels (artefacts) will be added to the final segmented image. Hence, we have used the method presented in reference \cite{soares2006retinal} to cope with this issue. An example of how this method works is illustrated in Fig. \ref{boundary-expansion-image}. As can be seen in the figure, the pixels close to the retina border, on the other side of the border (of the green line), are a replication of the pixels in the inner part of the retina. This largely eliminates the edge-like structure of the borders of the retina.

In the following sections, we introduce machine learning and optimization tools used to eliminate bright lesions from retina images. How these methods work is discussed in Chapter 4. 

\section{K-Means Segmentation}

In machine learning, k-means clustering is an iterative method of clustering data samples (here separating image pixels with different intensities), so that each sample fits into a cluster that has the closest mean to that sample. Samples that are close to a cluster gain more weight and thereby eliminate the effects of dissimilar ones \cite{zhou2010digital}. 

Depending on the value of the intensity, the retina pixels can be grouped into three categories (clusters): dark pixels (vessels), foreground or bright pixels (such as optical discs and bright lesions), and background pixels. The background pixels here are those that are not located on vessels, bright lesions, and the optical disk, and their intensities range between dark and bright. Based on what has been said, the steps of the k-means clustering algorithm will be as follows (this algorithm was proposed for two clusters and for separating background pixels from the vessel pixels in reference \cite{aibinu2010vascular}. In this thesis an extended version of this algorithm is utilized):

\begin{itemize}
    \item[$Step 1:$]  Select the number of clusters (here 3), and compute the initial value \(c_v(0)\), \(c_b(0)\), and \(c_f(0)\), as the mean values for vessel pixels, the background, and the foreground pixels, respectively:
    \[c_v(0) = m - \sigma, c_b(0) = m, c_v(0) = m + \sigma\]
where \(m\) and \(\sigma\) are the mean and standard deviation of the intensity matrix of the retina field of view computed as follows:
\[m = \frac{\sum I_p}{N}, \sigma = \sqrt{\frac{\sum(I_p - m)^2}{N}}\]
in which \(I_p\) is the pixel intensity of pixel \(p\) in the image and \(N\) is the number of pixels in the field of view. 
    \item[$Step 2:$] Find the distance of each pixel's intensity with the mean of each cluster and divide the image's pixels into three groups based on the minimum distance from the mean of the cluster.
    \[I_i^{(n)} = \{I_p:||I_p-c_i^{(n)}||=min_j||I_p-c_j^{(n)}|| \forall i,j \in \{v, b, c\}\}\]
In which, \(I_p\) is the intensity of the pixel \(p\) in the image and \(n\) is the current iteration number of the algorithm. \(I_i^{(n)}\) is the set of pixels that fall into the cluster \(i\) during \(n\)th iteration. 
    \item[$Step 3:$]  Compute the new mean for each cluster with the following formulas:
    \[c_v^{(n+1)} = \frac{\sum I_v^{(n)}}{N_v}, c_b^{(n+1)} = \frac{\sum I_b^{(n)}}{N_b}, c_f^{(n+1)} = \frac{\sum I_f^{(n)}}{N_f} \]
Where \(I_v^{(n)}\), \(I_b^{(n)}\), \(I_f^{(n)}\) are three clusters (a group of pixels) in \(n\)th iteration, in which the intensities of the vascular, background, and foreground pixels, respectively, are placed.

    \item[$Step 4:$]  If each of the mean values is changed, put \(n=n+1\), and go back to Step 2, otherwise go to Step 5.
    \item[$Step 5:$] Finish.
\end{itemize}

In Chapter 4, we will see how the mean obtained for the foreground cluster, \(c_f\), will be used to eliminate bright lesions.
\section{Structurally Differentiable (SD) Plane (SDP)}
The structurally differentiable plane is a plane in which portions of the input image that have sharp changes (sharp fluctuations) in intensity, such as the edges of bright lesions, have disappeared, and edge-like structures whose changes in intensity are mild, such as vessels, are more signified. In fact, the SD plane responds small to discontinued and noisy signals and responds big to smoother signals. To produce this plane, we use a regularization procedure.

The SDP is obtained by generating the negation (inverted) plane and subtracting that from the input image. An inverted plane is a plane with a contrasting meaning to the SDP. On this plane, unlike the SDP, noisy structures, including bright lesions in retinal images, and softer structures, such as vessels, are more smooth. The inverted plane can be obtained by minimizing the "Total Variations" model (energy). The Total Variations model is given by the following relationship \cite{lam2010general}, \cite{shen2002weber}; a more detailed description of the model is in Appendix B: 
\[E(v) = \int_\Omega \{||\nabla v||+\frac{\lambda}{2}\int_\Omega ||v^0-v^{SD}||^2\}dxdy\]
Which \(v^0\) is the input image, \(v^{SD}\) is the negation of SDP and \(||\nabla v||_\tau = \sqrt{\tau^2+|\nabla_v|^2}\). The parameter \(\tau\) is only for computational needs and to prevent zero magnitudes for the gradient, \(||\nabla v||\) where it comes under the fraction. If the size of this parameter is small, it does not affect the optimization response. Therefore, in all experiments of this study and after examining its effect on the segmentation, its value is very small and equal \(10^{-2}\).

\(\lambda\) is the Lagrangian fitness or coefficient parameter that is chosen to establish one of the optimization constraints. The value of this parameter is important for balancing image smoothing and eliminating bright lesions. A more detailed explanation of how to obtain these parameters is provided in Appendix B.

By applying the calculus of variations, such as differential equations, to compute the minimum value for \(E\), the following Euler-Lagrange equation is obtained in which the optimized value of \(v^{SD}\) holds.

\begin{equation} \label{formula-3_2}
\nabla.\frac{\nabla v^{SD}}{||\nabla v^{SD}||_2} + \lambda (v^0 - v^{SD}) = 0
\end{equation}

The optimal value for \(v^{SD}\) is obtained by the Digital Total Variation (TV) filter. Appendix B shows how to apply this filter. 

Regularization obtains the degree of the concavity (rate of change in the magnitude of the intensity) in the plane \(v^{SD}\). According to \ref{formula-3_2}, if the intensity difference between a pixel and its neighbours is high, \(\frac{1}{||\nabla v^{SD}||_\tau}\) will become very small, and the regularization term, \(\nabla.\frac{\nabla v^{SD}}{||\nabla v^{SD}||_\tau}\) at this pixel will greatly reduce, approaching zero, and we will have: 
\[\lambda(v^0-v^{SD})=0\]
So, the intensity of this pixel on the plane \(v^{SD}\) will become equal to its value on the plane \(v^0\) (the input image).

\begin{figure}[ht!]
\begin{center}
\includegraphics[scale=0.5]{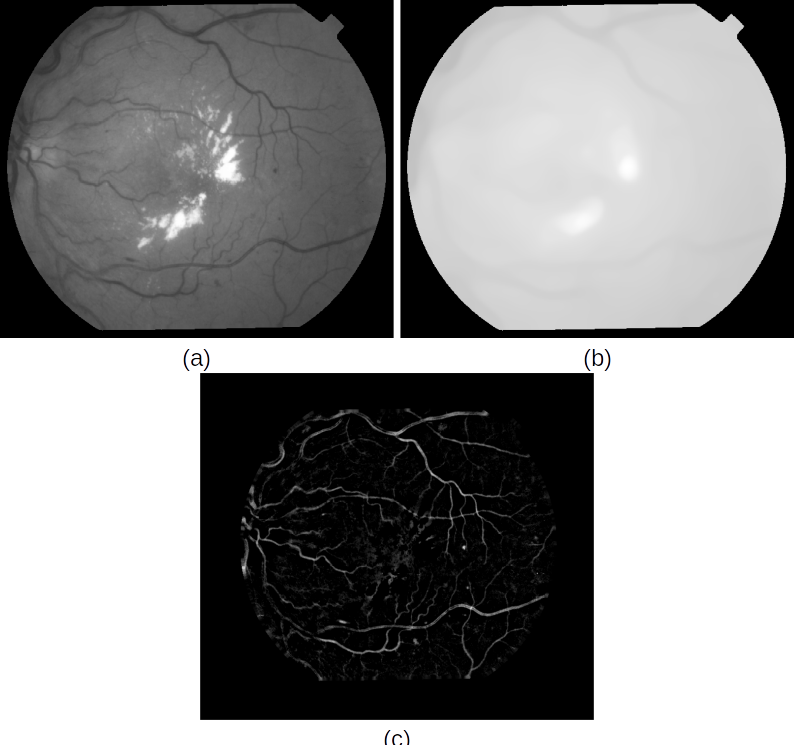}
\end{center}
\caption{Using regularization to reduce bright lesions. (a) Input image. (b) Regularizaiton output, \(v^{SD}\). (c) The structurally differentiable plane, \(C_SD\)}
%Source:
\label{image-regularization-bright-lesion}
\end{figure}

The pattern of intensity change of bright lesions has a steep slope, but the vessels have a pattern of smooth intensity change. Because of this smoothness, the value of the regularization term in the pixel of the vessels is much greater than that for the bright lesions with high roughness. Part (b) of Fig. \ref{image-regularization-bright-lesion} shows the output of the Total Variation model. 

In this image, bright lesions are seen in the middle of the image, and other portions, especially concavities (such as vessels), are largely smoothed out. By subtracting the input image \(v^0\) (part (a) in Fig. \ref{image-regularization-bright-lesion}) from the plane \(v^{SD}\), the SD plane, \(C_{SD}=v^{SD}-v^0\), is obtained. When \(v^0\) is large (bright lesions), \(C_{SD}\) returns a small value, and vice versa (part (c) of Fig. \ref{image-regularization-bright-lesion}). In Chapter 4, the effect of this method on improving the outcome of vessel extraction is examined.

After examining the processes of reducing and eliminating bright lesions, in the sequel of this section, we introduce the line operator used in the proposed vessel extraction methods in Chapter 4. 
\section{Line Operator}
As mentioned earlier, the distinctive feature of the vessels is their linearity. Many methods have been proposed for line detection in images, such as the Hough Transform \cite{gonzales2002digital}, Oriented Bins line operator, Gaussian Derivatives, and ridge detectors. In \cite{zwiggelaar2004linear}, the efficacy of various methods for identifying linear vessel structures in mammographic images has been compared. The results showed that the line operator (and another method) provided the best performance in terms of signal-to-noise ratio, the accuracy of positioning the vessels, their alignment and their width.

In this thesis, we seek a linear detector that, in addition to performance, has the least number of parameters to be adjusted; its response time is short; and because the vessels have different widths and conditions, it can be applied in a multi-scale fashion. Therefore, the line operator method was chosen to identify the linear structures of the retinal vessels. In addition to being able to perform linear structure detection in a multi-scale process, this operator requires little time to respond. Also, the only parameter that needs to be adjusted is the window size, which in this study (Chapter 4), three sizes are chosen, to address single-operator deficiencies, improve reliability, and extend the detection method to detect vessels in images with different sizes.

In order to obtain an efficient line detection method, in addition to the line operator, the Hough Transform and the proposed method in \cite{liu1999line} were implemented. The former due to the lack of desirable response and the later due to the need for very high response time are rejected.

\begin{figure}[ht!]
\begin{center}
\includegraphics[scale=0.5]{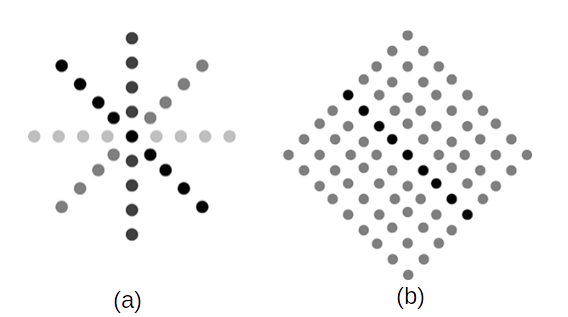}
\end{center}
\caption{Line operator. (a) The average intensity value, \(L_\theta\) of the pixels along the lines in several directions, \(\theta\). (b) The line response, \(R\), is the difference between the winner line, \(L_w\) (the darker pixels), and the average intensity of the parallel lines to the winner line in its squred neighbourhood, \(N_w\) (the brighter pixels): \(R=L_w-N_w\)}
%Source:
\label{3-7-image-Line-Operation}
\end{figure}

How the line operator works is shown in Fig. \ref{3-7-image-Line-Operation} . The average intensity, \(L_\theta\), between pixels on a line passing through a given pixel, is obtained in several directions, \(\theta\), and the direction at which the average is the highest is selected as the winner line, \(L_w\). To differentiate pixels on linear structures from other pixels, a criterion called Line Response, \(R\), is introduced. The line response of a pixel is obtained by the formula, \(R=L_w-N_w\), where \(N_w\) is the average intensity in a squared neighbourhood, in the same direction as the winning line (Part (b) in Fig. \ref{3-7-image-Line-Operation}).

The average line and its mean square neighbourhood are provided by passing a window through the image. This is a combination of Bi-Linear Interpolation in two directions (vertical and horizontal), sampling and averaging. 

The idea behind this process is that if the pixel at the centre of the intersection of the lines is a vessel pixel where the winning line is placed on a vessel, the response of the line will be large. On the other hand, the difference between \(L_w\) and \(N_w\) for the background pixels, which have a near-uniform neighbourhood, is very small and the line response for this part will be much smaller than that of vascular structures.

Chapter 4 explores why and how to apply this line operator in a multidimensional fashion and how it functions in the process of vessel detection in retinal images. In the methods of these thesis, after the line detection is completed, a threshold is applied to obtain the binary image, which we now examine how the threshold value is obtained.

\section{Thresholding}
For the output of the methods and for performance evaluation, after each segmentation method, a binary image of the output image is required, in which \(1\)'s represent the vessel pixels and \(0\)'s represent the non-vessel pixels. To obtain such an image, the output of the multi-scale line operator is thresholded. 

The threshold value was set such that the mean for “False-Negative rate” was 0.02. This value is determined by the fact that in medical applications it is best to minimize the number of pixels that are misdiagnosed in order to make the least error in identifying other parts. Therefore, the threshold value on the normalized output image was set 0.77 for method one and 1.25 for method two. The standard image is an image whose mean is zero and its variance is one. 

In this chapter, the processing tools used to implement the proposed methods of this thesis were presented. These tools include binary pattern acquisition, green channel image, preprocessing techniques such as transforming to Webber (perceived) and expanding boundaries, as well as algorithms used to remove anomalies and vessel detection, such as k-means clustering, line operator and thresholding. How to combine these tools to form the proposed methods for vessel extraction are presented in the next chapter.

\chapter{Blood Vessel Segmentation Methods}

In this chapter, the applied methods of this thesis for the automatic vessel extraction in retinal images are presented. In Chapter 3, the processing tools used in these methods are reviewed. In this chapter, we will discuss how to use these tools together to achieve effective methods, and also how these techniques work to reduce anomalies and detect vessels.

As illustrated in Chapter 1 (Fig. \ref{image-Methods}), in this thesis two different methods for detecting blood vessels in the retina image are presented. In both of these methods, first, a process is performed to reduce the impact of bright lesions. Then, a multidimensional line operator is applied to find the linear structures and to eliminate the dark anomalies with irregular structures. Henceforth, the method in which bright lesions have been reduced by means of the k-means clustering is called "Method 1", and the method in which the regularization is used is called "Method 2". In the remainder of this chapter, we first examine the step in the two methods in which the bright lesions are reduced. Subsequently, the multi-scale line operator, common between the two methods, is explained.

\section{Elimination of Bright Lesions}
\begin{figure}[ht!]
\begin{center}
\includegraphics[scale=0.5]{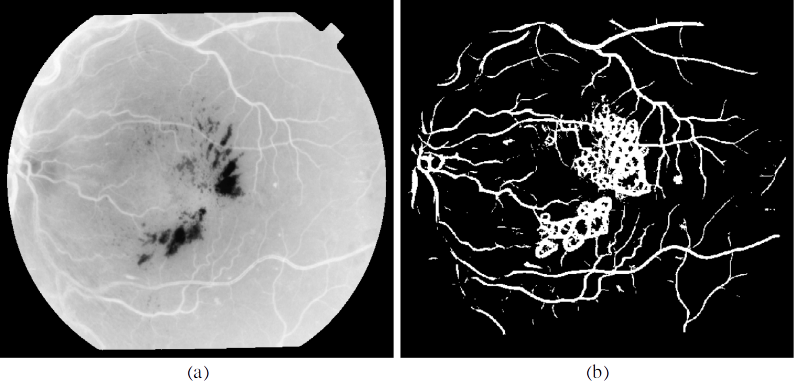}
\end{center}
\caption{The ringing effect. (a) The complement of the input image. (b) The result of vessel segmentation in the input image without removing the bright lesions and the occurrence of ringing effect}
%Source:
\label{image-ringing-effect}
\end{figure}

If no bright lesion in the patient's retina image is eliminated, the Ringing Effect occurs by detecting linear structures in the complement image input during the line detection algorithm. This happens due to the sharp and noise-like variations in bright lesions (Fig. \ref{image-ringing-effect}). In this case, when the bright lesions are filled (solid, no hole in it), the edges and pixels near it are inaccurately identified as vessels. When the line operator is used to detect vessels, since the size of the solid anomalies is larger than the size of the line operator windows, it gives small values to the pixels in the middle of it. However, at the edges, since the intensity difference of the pixels outside and within the bright lesions is large, the operator gives large values to the pixels near and on the edges. Here we present the methods to prevent this condition by using anomaly reduction procedures before the line operator.

\subsection{Reducing Bright Lesions in Method 1}
\begin{figure}[ht!]
\begin{center}
\includegraphics[scale=0.5]{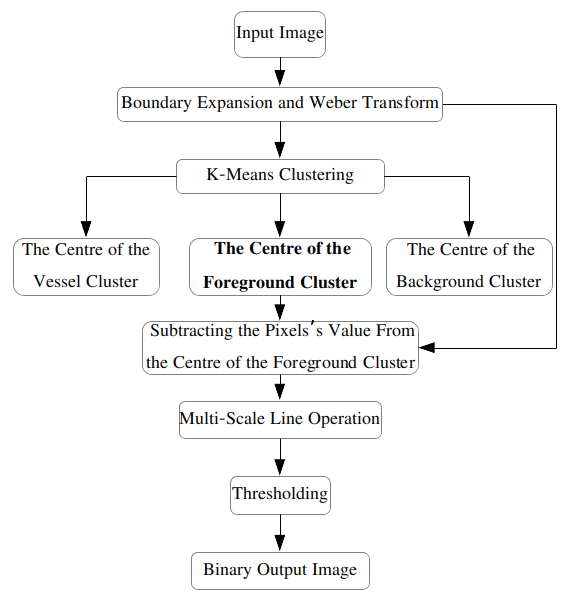}
\end{center}
\caption{The diagram of Method 1 - K-Means clustering and multi-scale line detection}
%Source:
\label{image-Method1}
\end{figure}

Fig. \ref{image-Method1} illustrates how to use a k-means clustering method before the multi-scale line operator to identify retinal vessels. In this approach, the k-means clustering method has a fundamental role in reducing bright lesions. This clustering approach is very simple and only based on the pixels' intensity values (grey level) feature. The clustering procedure is used to provide a plane in which the significance of bright lesions is reduced based on their intensity characteristics. 

In Section 3.4, the steps of the k-means algorithm are presented. Using this algorithm, the pixels of the image are clustered into three categories, vessels, background, and foreground, and the midpoints of the clusters are obtained. 

With these midpoint values, one can build three planes, where the value of each pixel equals the difference in intensity of that pixel in the input image with the midpoint (mean value) of each cluster. In fact, these planes represent the clustering response in a different way. In them, the value of each pixel is equal to the closeness of the intensity of that pixel to the middle of each cluster. 

Here, we only need a plane where the value of each pixel is obtained by subtracting the value in the input image from the middle of the foreground cluster. To do this, the plane is created as follows:
\begin{equation} \label{differentiable-formula}
df_p = c_f - I_p    
\end{equation}

Where \(I_p\) is the intensity of pixel \(p\) of the input image, and \(c_f\) is equal to the mean of the foreground cluster (bright pixels) and \(df_p\) (the distance from the middle of the foreground cluster) is the value of the pixel \(p\) on the plane \(df\).

Because \(c_f\) is the middle of a cluster that contains (most likely) optical disk pixels and bright lesions, it can be best estimated for the intensity of bright pixels. So if \(p\) is a vessel pixel, which is darker than other parts, \(df_p\) will have a high value, and if \(p\) is a background or foreground (bright) pixel, \(df_p\) will have a smaller value.

\begin{figure}[ht!]
\begin{center}
\includegraphics[scale=0.5]{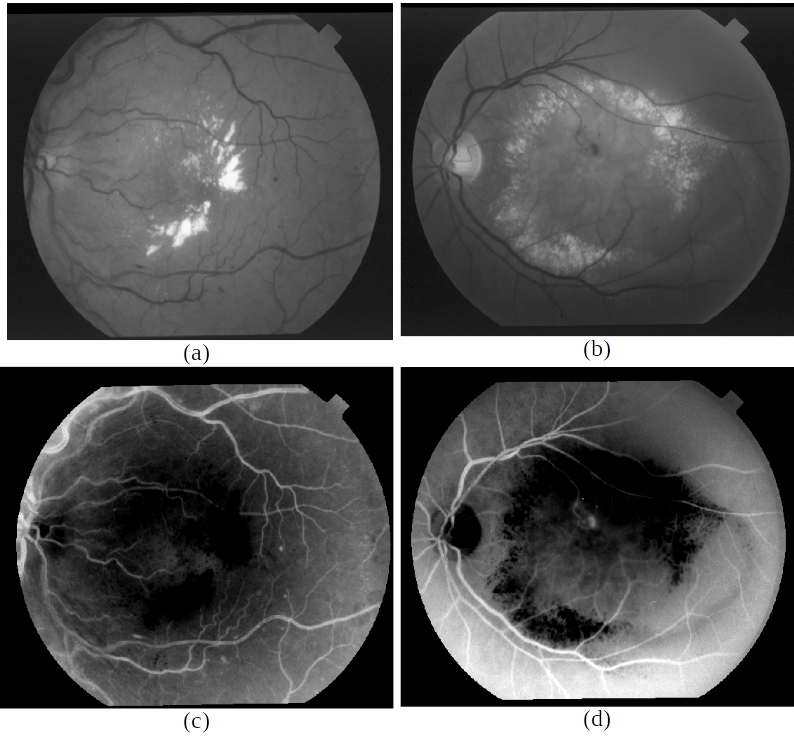}
\end{center}
\caption{The effect of k-means clustering on eliminating the bright lesions (images are scaled to [0,1]). (a),(b) are two samples of input images. (c),(d) are the corresponding plane \(df\) for (a) and (b) respectively.}
%Source:
\label{image-k-means-bright-lesions}
\end{figure}

The effect of applying the method described above to remove the bright lesions is shown in parts (c) and (d) of Fig. \ref{image-k-means-bright-lesions}. As can be seen in these images, the intensity difference of the areas with bright lesions with the background is greatly reduced, and the line detector will no longer be misled in the sections near the boundary of these anomalies. Sections with abnormalities have also gone out of being noise-shape and become more uniform. Figure 4.4 illustrates the effect of k-means clustering on improving vessel detection with the line operators.

\begin{figure}[ht!]
\begin{center}
\includegraphics[scale=0.5]{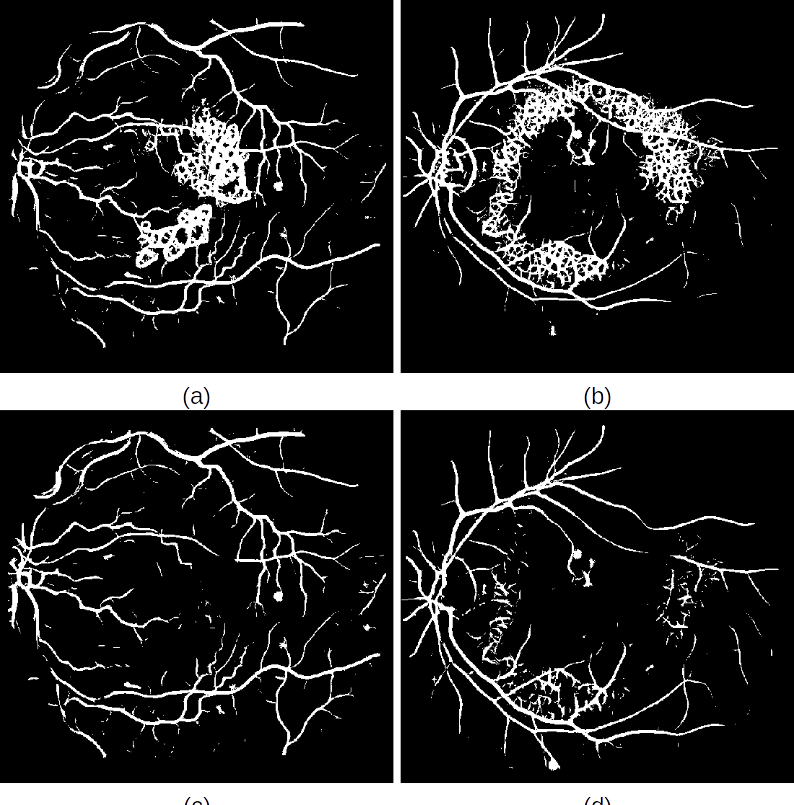}
\end{center}
\caption{The effect of k-means clustering on the final vessel segmentation. (a), (b) the ringing effect occurred in the final segmentation of parts (a) and (b) of Fig. \ref{image-k-means-bright-lesions}. (c) , (d) are the final results of vessel segmentation with eliminating of bright lesions using k-means clustering in parts (a) and (b) of Fig. \ref{image-k-means-bright-lesions}.}
%Source:
\label{image-k-means-bright-lesions-seg-result}
\end{figure}

\subsection{Reducing Bright Lesions in Method 2}
\begin{figure}[ht!]
\begin{center}
\includegraphics[scale=0.5]{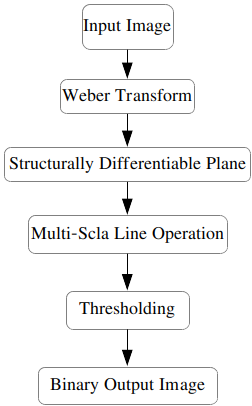}
\end{center}
\caption{The diagram of Method 2.}
%Source:
\label{image-diagram-Method2}
\end{figure}

Fig. \ref{image-diagram-Method2} shows how to use the SDP, before the line operator, to reduce bright lesions.

The pattern of the intensity values in the vessels is concave with a slight slope and in bright lesions, it is rough and fluctuates with a steep slope. Smoothing the whole image, while eliminating rough gradients, also smooths the vessels. Therefore, a regularization procedure can be used to maintain the structures with well-graded but gentle gradients (vessels). The SDP is the result of such regularization. 

\begin{figure}[ht!]
\begin{center}
\includegraphics[scale=.4]{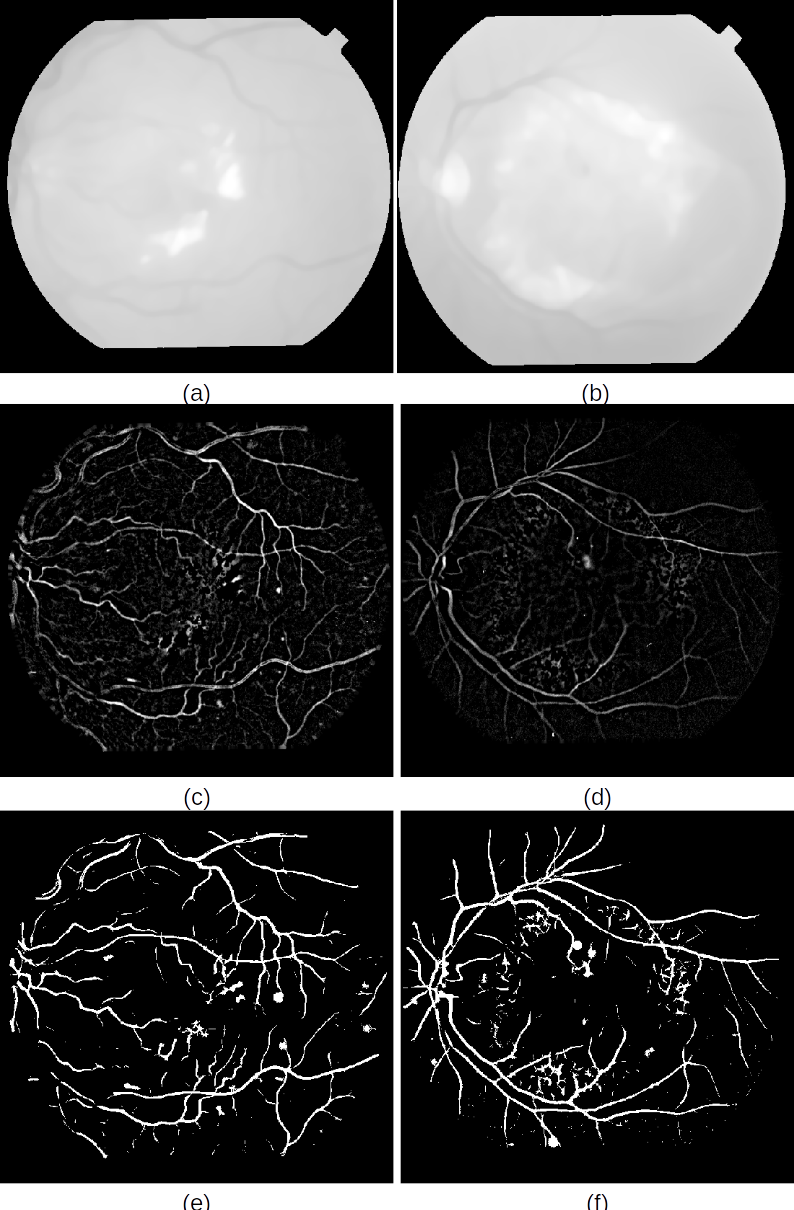}
\end{center}
\caption{Improving the performance of Linear structure segmentation using the structurally differentiable plane. (a), (b) are the inverted plane, \(c^{SD}\), of the images in Parts (a) and (b) of Fig. \ref{image-k-means-bright-lesions}. (c) and (d) are the SD planes of (a) and (b). (e) and (f) The vessel segmentation results.}
%Source:
\label{image-Differentiable-Plane-Result}
\end{figure}

As mentioned earlier, the intensity pattern of bright lesions makes their behaviour similar to that of noise. Therefore, in Method 2, a regularization method based on the noise removal Total Variation process is used. As explained in 3.7, the SD plane, \(C_{SD}\) is obtained by minimizing its inverted plane, \(v^{SD}\), from the input image, \(v^0\). Parts (a) and (b) in Fig. \ref{image-Differentiable-Plane-Result} are the inverted planes corresponding to the structurally differentiable planes in parts (a) and (b) in Fig. \ref{image-k-means-bright-lesions}. As can be seen, the bright lesions in these images are well preserved and the vessels are smoothed out. The SDP for these images is also shown in pars (c) and (d) of Fig. \ref{image-Differentiable-Plane-Result}. In these images, as expected, the vessels are significant and the abnormalities have been greatly weakened. Parts (e) and (f) of this figure illustrate the final vessel segmentation results, which are the result of the use of a multi-scale line operator on the SDP.

\section{Detection of Vessels with Multi-Scale line operator}
After examining the methods for reducing bright lesions, we now examine how to identify vessels; the multi-scale line operating algorithm. 
As stated in Chapter 3, in this thesis three scales for detection the linear structure of vessels are chosen. Lines (windows) with dimensions of 5 (5x5), 11 (11x11) and 15 (15x15) and 12 angles (15 degrees apart) are used to identify the line in three levels.

\begin{figure}[ht!]
\begin{center}
\includegraphics[scale=0.5]{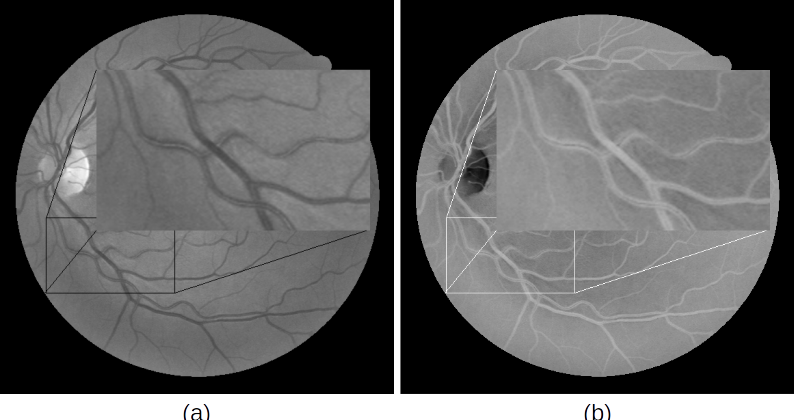}
\end{center}
\caption{The vessel centre light reflection. (a) This parts in the original image. (b) This pars in the inverted image.}
%Source:
\label{4.7-image-Central-Reflex}
\end{figure}

It has been shown \cite{ricci2007retinal} that line size 15 is a good choice for detecting retinal vessels. A line operator in this size is good at dealing with light reflection from the middle of the vessels. In this case, the intensity of the pixels in the middle of the vessels is less than the intensities of the surrounding pixels (mostly in the complement image) (Fig. \ref{4.7-image-Central-Reflex}). This leads to a confusion of the segmentation process and an increase in False-Positive responses. A line operator with size 15 can identify (highlight) these pixels because, in such a situation, the winning line only has a small number of pixels in the middle part of the vessel. Therefore, this low number of mid pixels cannot have a significant effect on the mean of the lines results and the response of the line will also be high. Also, since the size of window 15 is usually larger than the width of the vessels, most background pixels are correctly identified. 

On the other hand, this operator also has some drawbacks. One of the disadvantages is that due to the large size of the window, the pixels in the spaces between the two vessels may also be inaccurately identified when the two vessels are too close to each other (Part (a) in Fig. 4.8). This problem also occurs among the vessel intersections (Part (b) in Fig. 4.8). Also, this operator may misidentify the out-of-vessel pixels along with vessels whose intensity values are strong (Part (c) in Fig. 4.8).

The suggested solution for these deficiencies is to use multiple sizes for the line operator window, and then to aggregate the results obtained by all three scales. 

\begin{figure}[ht!]
\begin{center}
\includegraphics[scale=0.5]{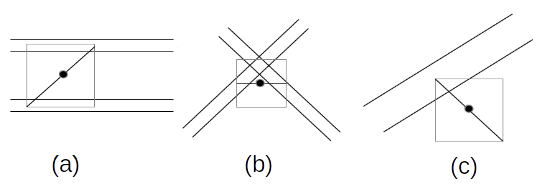}
\end{center}
\caption{The position of the line operator in which it operates incorrectly. (a) The pixels between two vessels. (b) Pixels close to vessels bifurcations. (c) Pixels near strong vessels.}
%Source:
\label{4.8-image-Line-Operator-Incorrect}
\end{figure}

\begin{figure}[ht!]
\begin{center}
\includegraphics[scale=0.5]{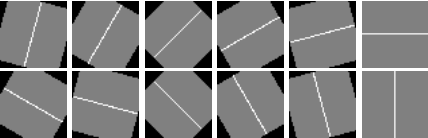}
\end{center}
\caption{The diagram of Method 2.}
%Source:
\label{1.9-image-lines-of-different-angles}
\end{figure}

In this study, by investigating the effect of using the size of lines between 3 and 15 pixels on vessel segmentation and considering response time and the quality of the response itself, it turned out that exploiting three-dimensional lines of sizes 5, 11, and 15 leads to an acceptable efficiency. The two lines of length 11 and 15 are good at dealing with vessel central reflection but have the defects investigated. These deficiencies are largely remedied by the use of a size 5 line, but as we will see later, this line generates background noises. Thus, combining the responses of all three lines can cover the shortcomings of each and increases the reliability of the obtained response. In Fig. \ref{1.9-image-lines-of-different-angles}, the sample windows for each of the twelve angles are shown.

To detect a line with a multi-scale line operator, first, all three windows \(5\times5\) (lines with length 5), \(11\times11\) (lines with length 11) and \(15\times15\) (lines with length 15) are passed through the image with the process explained in 3.8. The line response, \(R\), for each pixel, is obtained by each of these windows. In the end, the line response value for each pixel, \(R_{combination}\), equals the sum of all the responses of the three windows. \[R_{combination} = R_{5\times5}+ R_{11\times11}+{R_15\times15}\]
In this way, in addition to providing a relatively uniform condition for all vessels of varying widths using different line sizes, the individual failures of each line can be partially covered.

\begin{figure}[ht!]
\begin{center}
\includegraphics[scale=.4]{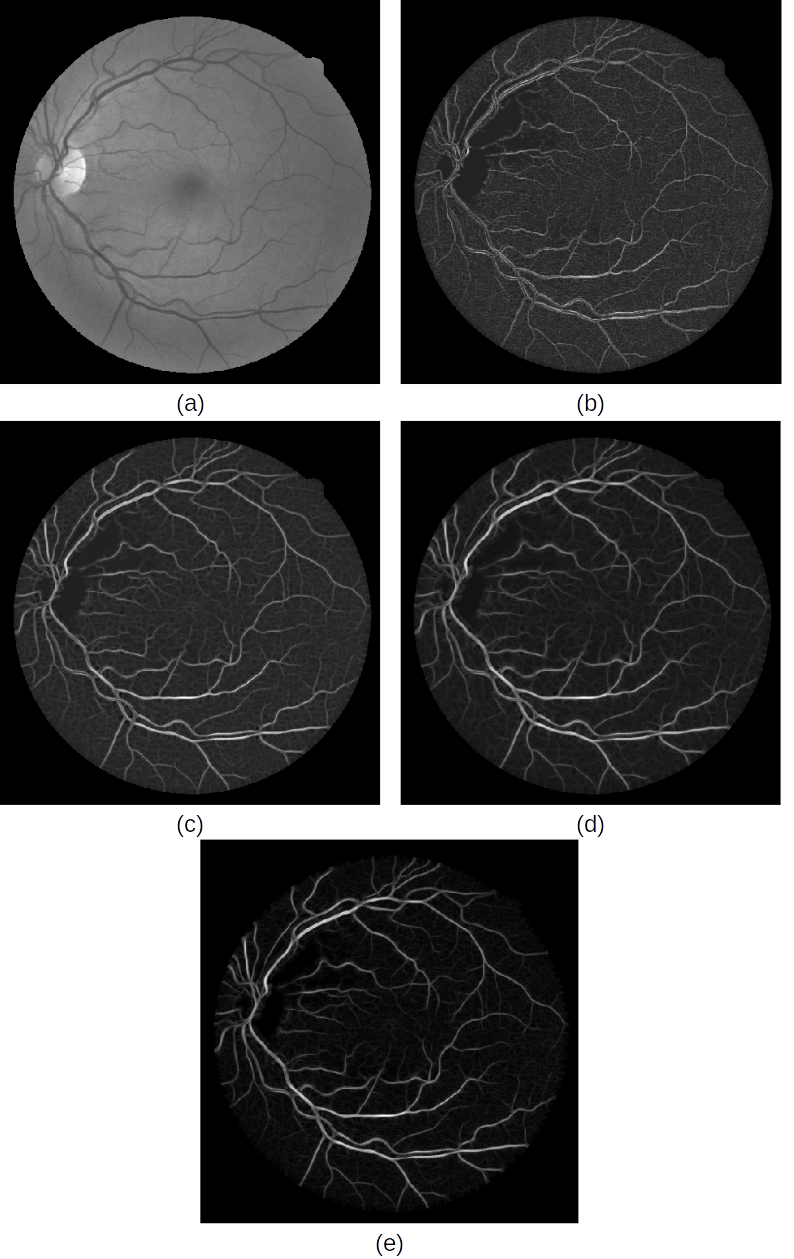}
\end{center}
\caption{The vessel segmentation result of line operator with different sizes. (a) The input image. (b) \(R_{5\times5}\). (b) \(R_{11\times11}\). (c) \(R_{15\times15}\)}
%Source:
\label{4.10-image-Linear-vessel-detection}
\end{figure}

Fig. \ref{4.10-image-Linear-vessel-detection} shows an example of the vessel segmentation process after the procedure of anomaly reduction in Method 2. As can be seen in this figure, by combining the responses of all three windows, the vessels are better highlighted and the result of segmentation becomes more accurate. 

Below we examine the function of the line operator in various parts of the retina image. Note that the input of the multi-scale line detector algorithm is the image in which the vessels are bright (the output of one of the anomaly reduction algorithms).

\subsection{How the Line Operator Works In the Retina Image}
As shown in Fig. \ref{4.10-image-Linear-vessel-detection}, the larger the line size, the better it appears the inside parts of the vessels. When the vessel is larger than the window, the entire window fits into the vessel. Therefore, the intensity of the pixels in the neighbourhood of the pixel on which the window is located is close together and the average difference (line response \(R\)) for that pixel becomes very small. In other words, short lines (small windows) deal with large vessels, as the background areas. The larger the line, the less likely the entire window will fit into the vessel.

In Part (b) of Fig. \ref{4.10-image-Linear-vessel-detection}, the \(5\times5\) window has failed to highlight well the parts within the large vessels. Also, because this window works in very small areas, the line response is large for many background pixels and has caused a noisy response. But this value is much smaller than the linear response given by larger windows for the same areas, therefore, it will disappear at the final thresholding.

As mentioned earlier, the larger the window of the line operators, the better the operator can highlight the vessels that have a central reflection. This is well illustrated in Parts (c) and (d) of figure \ref{4.10-image-Linear-vessel-detection}.

Retina images sometimes have dark anomalies, such as hemorrhages, which are similar to vessels intensity-wise. The pattern (texture) of the intensity of vessels is uniformly drawn in one direction, but the pattern of intensities of dark anomalies is non-uniformly distributed in one piece \cite{lam2010general}. These abnormalities are mostly round and small, which makes them different from linear structures. They can be partially ignored by using the line operator. The outcome of vessel detection along with these abnormalities is discussed further in chapter 5.

After the end of both extraction methods and after the multi-scale line operation, the thresholding procedure stated in 3.7 is used to obtain the binary output image.

In this chapter, the details of the vessel segmentation methods of this thesis were explained. First, the process of reducing bright lesions by means of k-means clustering in Method 1, and the structurally differentiable plane in Method 2 were investigated. Then, we delineated the method used for identification of linear structures with a multi-scale line operator. Then, how this operator copes with different parts of a retina image was discussed. In Chapter 5, the results of comparing the performance of the proposed methods with each other and with other methods are reported.

\chapter{Performance Evaluation}

After examining how the applied methods work in Chapter 4, we will evaluate their efficiency in this chapter. First, the criteria used for evaluation are described. Then, the efficacy of the methods is measured from the point of view of the accuracy of the vessel segmentation, for the patient and healthy retina images, as well as their response times. Finally, by providing some examples of the segmentation results the methods will be qualitatively analyzed. Examples of the best and worst answers will also be provided. 

\section{Evaluation Criteria}
What comes out at the end of the process of retinal vessel segmentation is a binary image. In this binary image, each pixel may belong to the vessel or non-vessel categories. Hence, four possible states are assumed for a classified pixel; two states for when it is correctly classified and two for the time that the classification process went wrong. The pixels in the first state are those that are either True-Positive or True-Negative; this happens when they have the same label (0 or 1) in both the labelled image and in the segmented image. On the other hand, the pixels in the second state are those that are either False-Positive or False-Negative; this occurs when there is a discrepancy between the labels of the pixel in the labelled image and in the segmented image.

The True-Positive rate indicates the ability of the method to find vessel pixels. However, False-Positive rate indicates how often the method confuses non-vessel pixels as vessel pixels. Accuracy equals the ratio of the number of pixels correctly identified (True-Positive and True-Negative pixels) to the number of all pixels in the image field of view. 

In many studies, the efficiency of segmentation methods is measured by a ROC diagram \cite{ricci2007retinal}. The ROC diagram represents the True-Positive rate versus the False-Positive rate. The higher the curve of the graph to the upper left corner, the better the performance of the system. What comes out of a ROC diagram to measure the efficiency of a method is the size of the area under the curve, which equals 1 for an optimal system \cite{fraz2012blood}. 

The mentioned measurements are estimated for the image field of view only. A summary of what has been explained is provided in Table 5.1 \cite{fraz2012blood}.

\begin{center}
\begin{table}[ht!]
\caption{The measurements used to evaluate the methods performance.}
\begin{tabular}{ | m{10em} | m{10cm}| } 
\hline
\textbf{Measure} & \textbf{Definition} \\ 
\hline
True-Positive Rate & \(\frac{Number of True-Positive Pixels}{Number of All Vessel Pixels}\) \\ 
\hline
False-Positive Rate & \(\frac{Number of False-Positive Pixels}{Number of All Non-Vessel Pixels}\) \\ 
\hline
Accuracy & \(\frac{Number of True-Negative Pixels + Number of True-Positive Pixels}{Number All the FOV Pixels}\) \\
\hline

\end{tabular}
%Source:

\label{5.1-Table-Measurements}
\end{table}
\end{center}

\section{Statistical Performance Evaluation}
Here we review the results obtained from the segmentation methods explained in Chapter 4 and compare them with other researches. All implementations were performed in MATLAB, on a laptop with a dual-core Duo (TM) processor CPU running at 2.24 GHz and 1 GB of RAM on  Windows 7.

In addition to adjusting the size of the lines in the line operator, the parameter \( \tau \) of the regularization method was set to \(10^{-4}\), and the number of iterations of the Total Variation filtering algorithm was set to \(100\). Higher values for iteration numbers also were tested. As the number of iterations increases, although the number of False-Positive pixels decreases, especially in bright lesions in abnormal retinal images, many True-Positive pixels disappear. The computing time also increases dramatically. Hence, the choice of the number of iterations, in addition to the efficiency, provided a good response time. As shown in Figure 5.1, the anomaly decreases with the increase in the number of iterations, but at the same time, more details of the image are eliminated.

\begin{figure}[ht!]
\begin{center}
\includegraphics[scale=0.5]{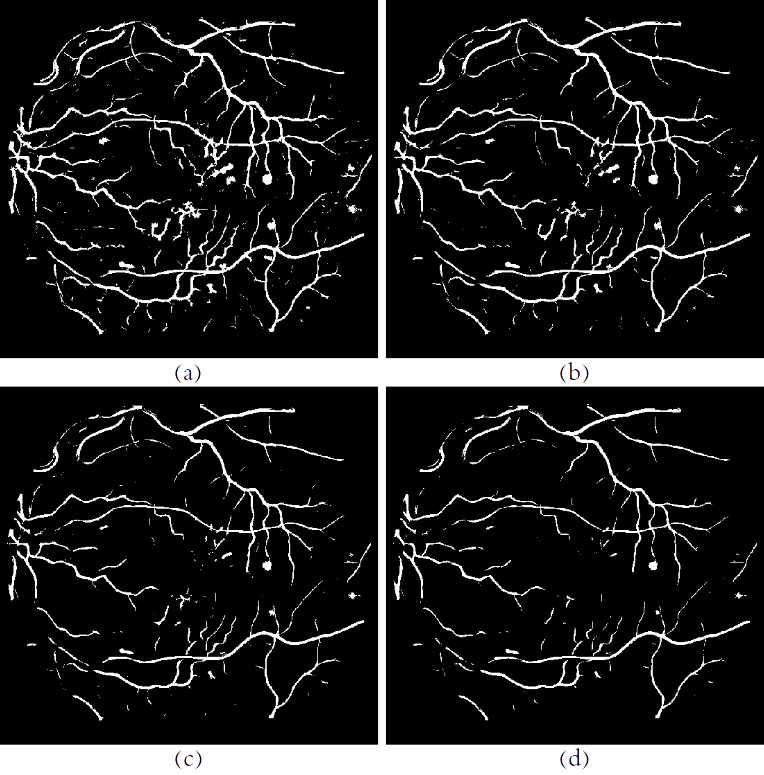}
\end{center}
\caption{The effect of increasing the iterations of the TV filter algorithm. (a) 100, (b) 200, (c) 300 and (d) 400 iterations.}
%Source:
\label{5.1-image-TV-Iterations-Effect}
\end{figure}

Previously, in Chapter 3, we discussed the parameter \(\lambda\). In Appendix B the calculation process is provided in more detail.

In 3.1, the image datasets used in this thesis are explained. As it turned out, each image in the STARE dataset has two labelled images, segmented by two experts: viewer 1, viewer 2. For images of this dataset, viewer 1 has marked 10.4 per cent and viewer 2 has marked 14.9 per cent of pixels as vessel pixels. The performance evaluation of the segmentation performed on the basis of viewer 1 notation.

The DRIVE dataset also has two series of labelled images: Category A and Category B. In group A, 12.7 per cent and in group B 12.3 per cent of pixels are categorized as vessel pixels. The efficiency of the segmentation methods on this set is based on Category B \cite{ricci2007retinal}.

In the remainder of this chapter, the results of the two methods will be measured together with all other images of these two sets, as well as with the pathological images (with anomalies) of the STARE dataset. The results will be compared with methods in other researches. The accuracy obtained for each image category (DRIVE, STARE, and pathological STARE images) is equal to the average accuracy of the vessel detection method in all images in that category and based on the thresholding method described in 3.7.
\subsection{Statistical Performance Evaluation of the Two Methods}
Figures 5.2, 5.3 and 5.4 illustrate the diagrams of the ROC obtained on all three categories of images. The continuous line shows the performance of Method 1 and the dashed line indicates the performance of Method 2. These graphs are drawn by interpolating 121 points. These points were obtained from the normalized responses of the thresholding method with threshold values in the range of -2 to 4, and averaging over True-Positive and False-Positive rates for each threshold value.

The "black dot" represents the average viewer 2 performance for the STARE and the Category B for the DRIVE dataset. In other words, the performance of vessel segmentation by viewer 2 in STARE and in Category B in DRIVE are evaluated based on the labelled images of viewer 1 and Category A, respectively. The average False-Positive rate for all STARE images is 0.061 for the viewer 2 and 0.903 for the True-Positive, and these averages for the pathological (abnormal) images are 0.04 and 0.81, respectively. For the DRIVE set, these values are 0.0275 and 0.775 for Category B, respectively.

\begin{figure}[ht!]
\begin{center}
\includegraphics[scale=0.5]{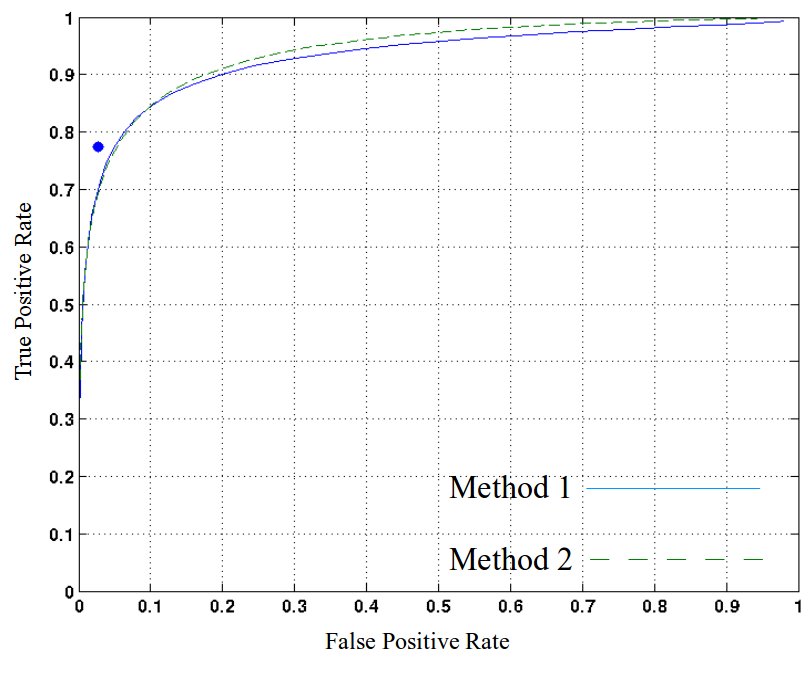}
\end{center}
\caption{The ROC curve relating to the vessel segmentation in images of DRIVE dataset for Method 1 and 2. The black dot is illustrates the True-Positive and False-Positive rates for the labelled images of Category B.}
%Source:
\label{5.2-ROC1}
\end{figure}

The ROC diagram, in Fig \ref{5.2-ROC1}, indicates the efficiency of the methods in segmenting vessels from DRIVE images. As can be seen, the breaking point of both methods occurred at a point with a True-Positive rate close to 0.75. That is, from this point on, with a slight increase in the True-Positive rate, the False-Positive rate will increase greatly, and vice versa. This indicates that the threshold values specified for the methods in Section 3.9 give the most reasonable output. 

On the other hand, as the True-Positive rate increases (as well as the False-Positive rate), the diagram of Method 2 rises above the diagram of Method 1. That is, where both methods have the same False-Positive rate, the True-Positive rate of Method 2 is greater. This shows that Method 2 works better than Method 1 in detecting thinner vessels. In other words, it extracts more details. In the qualitative analysis of the two methods, in 5.4, we will also see how Method 1 neglects some details. Also, where the False-Positive rate of Category B's (dark dot) labelled images is the same as our methods, the True-Positive rate of both methods is close to 0.7 and the True-Positive rate of the labelled image is close to 0.8. This shows that these methods require a 10 per cent improvement in the True-Positive rate.

\begin{figure}[ht!]
\begin{center}
\includegraphics[scale=0.5]{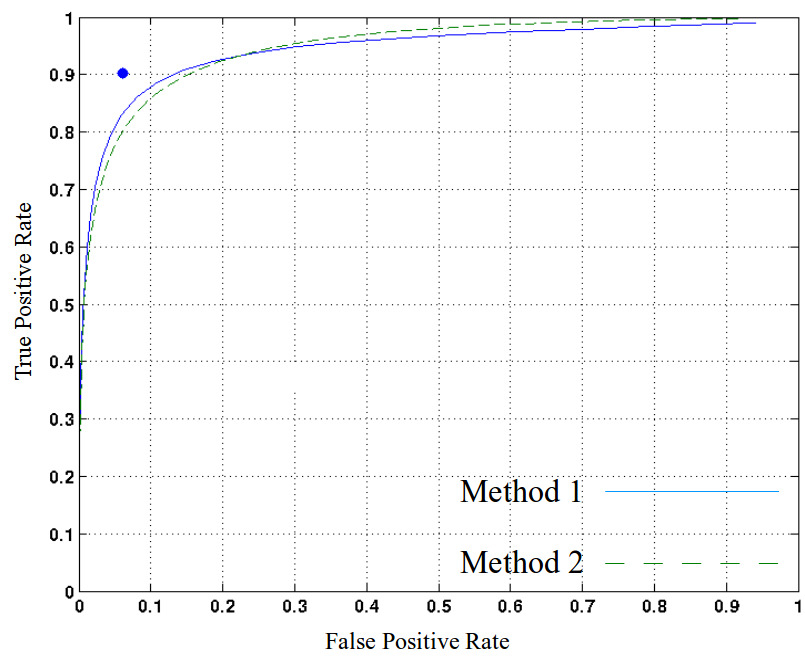}
\end{center}
\caption{The ROC curve relating to the vessel segmentation in images of STARE dataset for Method 1 and 2. The black dot is illustrates the True-Positive and False-Positive rates for the labelled images of the second reviewer.}
%Source:
\label{5.3-ROC1}
\end{figure}

In Fig. \ref{5.3-ROC1}, the performance diagram of the two methods for vessel segmentation in the STARE dataset is shown. Here, the curves break at a point where the True-Positive rate is close to 0.80. Also, the point at which the False-Positive rates of methods and viewers are equal shows that Method 1 and Method 2 need to improve their True-Positive rates by 5 per cent and 10 per cent, respectively. Here, too, one can see the same trend of a rising in curve of Method 2 than the curve of Method 1, when increasing the True-Positive rate. 

\begin{figure}[ht!]
\begin{center}
\includegraphics[scale=0.5]{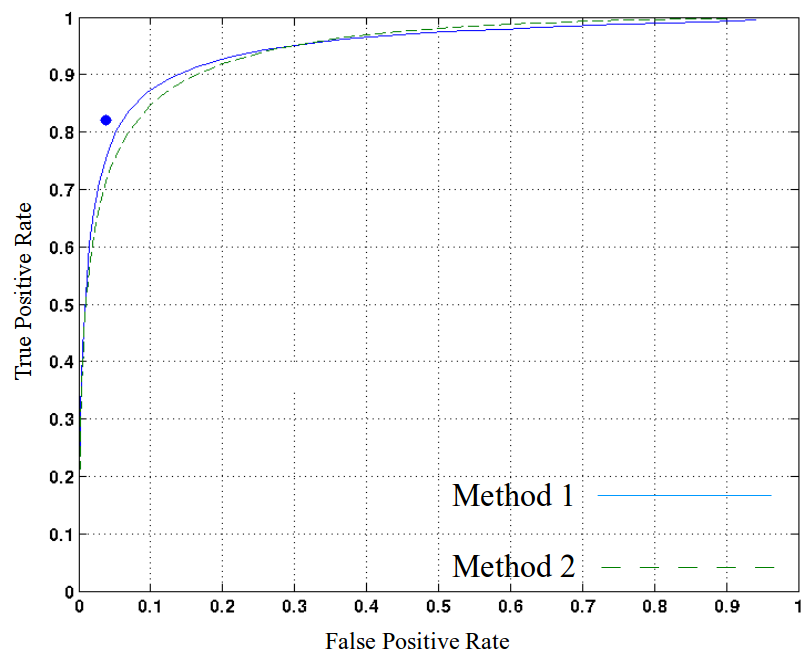}
\end{center}
\caption{The ROC curve relating to the vessel segmentation in pathological images of STARE dataset for Method 1 and 2. The black dot is illustrates the True-Positive and False-Positive rates for the labelled images of Category B.}
%Source:
\label{5.4-ROC1}
\end{figure}

The efficacy of the methods in detecting the retinal vessel pathological images in the STARE dataset is illustrated using the ROC diagram in Figure \ref{5.4-ROC1}. Breakpoints in this figure also occurred at the 0.75 True-Positive rates. Here, as in the previous two diagrams, after the breakpoint, the True-Positive rate of Method 2 has increased more than Method 1. Also, where the False-Positive rate of the methods and labelled images are equal, the True-Positive rate of the methods is approximately 10 per cent (slightly higher for Method 2 and lower for Method 1) than the True-Positive rate of the labelled images. 

As we have seen, ROC diagrams have had a relatively uniform overall trend for evaluating the results of the methods on all three image categories. This may indicate that both approaches exhibit uniform performance in relatively different conditions, without having to change their parameters. In the next chapter, we compare the two methods in this thesis with the other methods in the literature.
\subsection{Statistical Performance Evaluation with Previous Works}
Tables \ref{5.2-Table-DRIVE-Comparison-Accuracy} to \ref{5.7-Table-DRIVE-Comparison-Accuracy} show the statistics obtained from Methods 1 and 2, and the most referenced supervised and unsupervised methods in the literature, for segmentation accuracy and the area under the ROC curve.

\begin{center}
\begin{table}[ht!]
    \small
    \centering
    \caption{The Accuracy for The Vessel Segmentation in DRIVE Dataset}
    \begin{tabular}{|l|c|}
        \hline
        \multicolumn{1}{|c|}{\textbf{Method}} & \multicolumn{1}{|c|}{\textbf{Accuracy}} \\ 
        \hline
        \multicolumn{2}{|c|}{Unsupervised Methods} \\
        \hline
        Lam et al. \cite{lam2010general} & 0.9472\\
        \hline
         Ricci and Perfetti \cite{ricci2007retinal}(Implemented in \cite{lam2010general} & 0.9320 \\
        \hline
        Mendonca and Campilho \cite{mendonca2006segmentation} & 0.9463 \\
        \hline
        Jiang and Mojon \cite{jiang2003adaptive} & 0.8911 \\
        \hline
        Nguyen et al. \cite{nguyen2013effective}&0.9407\\
        \hline
        Method 1 & 0.9387\\
        \hline
        Method 2 & 0.9379\\
        \hline
        \multicolumn{2}{|c|}{Supervised Methods} \\
        \hline
        Niemeijer et al. \cite{niemeijer2004comparative}&0.9416\\
        \hline
        Staal et al. \cite{staal2004ridge} & 0.9441\\
        \hline
        Soares et al. (DRIVE) \cite{soares2006retinal} & 0.9466\\
        \hline
        Soares et al. (STARE) \cite{soares2006retinal} & 0.9445\\
        \hline
        Category B & 0.9473\\
        \hline
    \end{tabular}
    \label{5.2-Table-DRIVE-Comparison-Accuracy}
\end{table}
\end{center}

\begin{center}
\begin{table}[ht!]
    \small
    \centering
    \caption{The Area Under the Curve of the ROC curve for The Vessel Segmentation in DRIVE Dataset}
    \begin{tabular}{|l|c|}
        \hline
        \multicolumn{1}{|c|}{\textbf{Method}} & \multicolumn{1}{|c|}{\textbf{Accuracy}} \\
        \hline
        \multicolumn{2}{|c|}{Unsupervised Methods} \\
        \hline
        Lam et al. \cite{lam2010general} & 0.9614\\
        \hline
         Ricci and Perfetti \cite{ricci2007retinal}(Implemented in \cite{lam2010general} & 0.9410 \\
        \hline
        Jiang and Mojon \cite{jiang2003adaptive} & 0.0.9327 \\
        \hline
        Method 1 & 0.9303\\
        \hline
        Method 2 & 0.91413\\
        \hline
        \multicolumn{2}{|c|}{Supervised Methods} \\
        \hline
        Niemeijer et al. \cite{niemeijer2004comparative}&0.9294\\
        \hline
        Staal et al. \cite{staal2004ridge} & 0.9520\\
        \hline
        Soares et al. (DRIVE) \cite{soares2006retinal} & 0.9614\\
        \hline
        Soares et al. (STARE) \cite{soares2006retinal} & 0.9494\\
        \hline
    \end{tabular}
    \label{5.3-Table-DRIVE-Comparison-ROC-Area}
\end{table}
\end{center}

\begin{center}
\begin{table}[ht!]
    \small
    \centering
    \caption{The Accuracy for The Vessel Segmentation in STARE Dataset}
    \begin{tabular}{|l|c|}
        \hline
       \multicolumn{1}{|c|}{\textbf{Method}} & \multicolumn{1}{|c|}{\textbf{Accuracy}} \\
        \hline
        \multicolumn{2}{|c|}{Unsupervised Methods} \\
        \hline
        Lam et al. \cite{lam2010general} & 0.9567\\
        \hline
        Hoover et al. \cite{hoover2000locating}&0.9275\\
        \hline
         Ricci and Perfetti \cite{ricci2007retinal}(Implemented in \cite{lam2010general} & 0.9422 \\
        \hline
        Mendonca and Campilho \cite{mendonca2006segmentation} & 0.9479 \\
        \hline
        Jiang and Mojon \cite{jiang2003adaptive} & 0.9009 \\
        \hline
        Nguyen et al. \cite{nguyen2013effective}&0.9324\\
        \hline
        Method 1 & 0.9483\\
        \hline
        Method 2 & 0.9439\\
        \hline
        \multicolumn{2}{|c|}{Supervised Methods} \\
        \hline
        Staal et al. \cite{staal2004ridge} & 0.9516\\
        \hline
        Soares et al. (DRIVE) \cite{soares2006retinal} & 0.9469\\
        \hline
        Soares et al. (STARE) \cite{soares2006retinal} & 0.9480\\
        \hline
        Reviewer 2 & 0.9351\\
        \hline
    \end{tabular}
    \label{5.4-Table-DRIVE-Comparison-Accuracy}
\end{table}
\end{center}

\begin{center}
\begin{table}[ht!]
    \small
    \centering
    \caption{The Area Under the ROC Curve for The Vessel Segmentation in STARE Dataset}
    \begin{tabular}{|l|c|}
        \hline
        \multicolumn{1}{|c|}{\textbf{Method}} & \multicolumn{1}{|c|}{\textbf{Accuracy}} \\ 
        \hline
        \multicolumn{2}{|c|}{Unsupervised Methods} \\
        \hline
        Lam et al. \cite{lam2010general} & 0.9739\\
        \hline
        Hoover et al. \cite{hoover2000locating}&0.7590\\
        \hline
         Ricci and Perfetti \cite{ricci2007retinal}(Implemented in \cite{lam2010general} & 0.9615 \\
        \hline
        Jiang and Mojon \cite{jiang2003adaptive} & 0.9298 \\
        \hline
        Method 1 & 0.9431\\
        \hline
        Method 2 & 0.9471\\
        \hline
        \multicolumn{2}{|c|}{Supervised Methods} \\
        \hline
        Staal et al. \cite{staal2004ridge} & 0.9614\\
        \hline
        Soares et al. (DRIVE) \cite{soares2006retinal} & 0.9629\\
        \hline
        Soares et al. (STARE) \cite{soares2006retinal} & 0.9671\\
        \hline
        Reviewer 2 & 0.9351\\
        \hline
    \end{tabular}
    \label{5.5-Table-DRIVE-Comparison-Accuracy}
\end{table}
\end{center}

\begin{center}
\begin{table}[ht!]
    \small
    \centering
    \caption{The Accuracy for The Vessel Segmentation in Pathological Images of STARE Dataset}
    \begin{tabular}{|l|c|}
        \hline
       \multicolumn{1}{|c|}{\textbf{Method}} & \multicolumn{1}{|c|}{\textbf{Accuracy}} \\
        \hline
        \multicolumn{2}{|c|}{Unsupervised Methods} \\
        \hline
        Lam et al. \cite{lam2010general} & 0.9556\\
        \hline
         Ricci and Perfetti \cite{ricci2007retinal}(Implemented in \cite{lam2010general} & 0.9352 \\
        \hline
        Mendonca and Campilho \cite{mendonca2006segmentation} & 0.9426 \\
        \hline
        Jiang and Mojon \cite{jiang2003adaptive} & 0.9337 \\
        \hline
        Method 1 & 0.9438\\
        \hline
        Method 2 & 0.9404\\
        \hline
        \multicolumn{2}{|c|}{Supervised Methods} \\
        \hline
        Soares et al. (DRIVE) \cite{soares2006retinal} & 0.9428\\
        \hline
        Soares et al. (STARE) \cite{soares2006retinal} & 0.9425\\
        \hline
        Reviewer 2 & 0.9410\\
        \hline
    \end{tabular}
    \label{5.6-Table-DRIVE-Comparison-Accuracy}
\end{table}
\end{center}

\begin{center}
\begin{table}[ht!]
    \small
    \centering
    \caption{The Area Under the ROC Curve for The Vessel Segmentation in Pathological Images of STARE Dataset}
    \begin{tabular}{|l|c|}
        \hline
        \multicolumn{1}{|c|}{\textbf{Method}} & \multicolumn{1}{|c|}{\textbf{Accuracy}} \\
        \hline
        \multicolumn{2}{|c|}{Unsupervised Methods} \\
        \hline
        Lam et al. \cite{lam2010general} & 0.9707 \\
        \hline
         Ricci and Perfetti \cite{ricci2007retinal}(Implemented in \cite{lam2010general} & 0.9343 \\
        \hline
        Jiang and Mojon \cite{jiang2003adaptive} & 0.8906 \\
        \hline
        Method 1 & 0.9446\\
        \hline
        Method 2 & 0.9432\\
        \hline
        \multicolumn{2}{|c|}{Supervised Methods} \\
        \hline
        Soares et al. (DRIVE) \cite{soares2006retinal} & 0.9455\\
        \hline
        Soares et al. (STARE) \cite{soares2006retinal} & 0.9571\\
        \hline
    \end{tabular}
    \label{5.7-Table-DRIVE-Comparison-Accuracy}
\end{table}
\end{center}

The result of the supervised method described in \cite{soares2006retinal} is given once when training images were from the DRIVE dataset and once when these images were from the STARE dataset. Ricci and Perfetti's method  \cite{ding2011approach}, due to the use of the line operator, and Lam et al., because of the usage of two regularization procedures \cite{lam2010general} are the closest works to this thesis.

Looking at the tables, we find that the performance of both methods on the three image categories is closely comparable to state-of-the-art methods.

Overall, the performance of the two methods is very close to each other in terms of accuracy and area under the ROC curve. Their performance on DRIVE images is less than in STARE images. However, their performance on the pathological images of the STARE dataset is better than other methods and the second viewer.

Staal \cite{staal2004ridge} and Soares \cite{soares2006retinal} are better than Method 1 and 2, except in a few cases, but as reported in Staal \cite{staal2004ridge}, they employed only 19 images of the STARE dataset (9 pathological and 10 healthy).

Also the method of Nguyen et. al. \cite{nguyen2013effective} which uses all the line sizes (odd numbers) between 3 and 15, although in the DRIVE dataset is slightly better (from the accuracy perspective), but in the STARE dataset where the number of pathological (abnormal) images are more, is less effective than the proposed methods in this thesis. Furthermore, as the authors have argued, the threshold value of their method is determined to provide the highest average accuracy, albeit it is different from the threshold method described in this thesis, in Section 3.7, which was based on a low False-Negative rate.

The algorithms proposed by Ricci and Perfetti \cite{ricci2007retinal} and Lam \cite{lam2010general} are the best methods for the segmentation of vessels in healthy and pathological retinal images. Both Method 1 and Method 2 yielded close and sometimes better results than the Ricci and Perfetti's method, especially in the field of vessel detection in pathological images. However, the efficiency of the Lam’s method \cite{lam2010general} is always better than our methods. In addition to applying two simultaneous regularization procedures, a noise reduction method and a final optimization method have been applied in Lam's work, which increased the response time.

The response times of the two methods, along with the other methods, are shown in Table \ref{5.8-Running Time}. The average response time of the two methods was 7.6 seconds and 2.4 minutes, respectively. If the faster algorithms can be used to obtain the optimal regularization response and also in the line detection section, the response time can be reduced more. Implementation of methods in a programming language, such as C, can also help reducing response time.

\begin{center}
\begin{table}[ht!]
    \small
    \centering
    \caption{Running Time of Different Methods}
    \begin{tabular}{|m{4cm}|c|m{5cm}|}
        \hline
        \multicolumn{1}{|c|}{\textbf{Method}} & \multicolumn{1}{|c|}{\textbf{Running Time}} & \textbf{Computer Hardware}\\ 
        \hline
         Nguyen et al. \cite{nguyen2013effective} & 2.5 sec & Core 2 Duo Intel CPU, 2.4 GHz, 2 GB of RAM \\
        \hline
        Method 1 & 7.6 sec & Core 2 Duo (TM) Intel CPU, 2.24 GHz, 2 GB of RAM \\
        \hline
        Ricci and Perfetti  & 30 sec & Core Duo Intel CPU, 1.83 GHz, 2 GB of RAM \\
        \hline
        Method 2 & 2.4 min & Core 2 Duo (TM) Intel CPU, 2.24 GHz, 2 GB of RAM \\
        \hline
        Mendonca and Campilho \cite{mendonca2006segmentation}&2.5-3 min& Pentium 4 CPU, 3.2 GHz, 960 MB of RAM  \\
        \hline
        Soares et al.\cite{soares2006retinal}&3 min (9 hours training time)& PC wih 2167 Clock and 1 GB of RAM \\
        \hline
        Lam and Yan \cite{lam2008novel}&8 min&Duo 1.83 GHz CPU with 2 GB of RAM \\
        \hline
        Lam et al.\cite{lam2010general}&13 min&Duo 1.83 GHz CPU with 2 GB of RAM \\
        \hline
        Staal et al.\cite{staal2004ridge}&15 min&Pentium 4 1 GHz CPU \\
        \hline
    \end{tabular}
    \label{5.8-Running Time}
\end{table}
\end{center}
\section{Qualitative Evaluation}
Here are some examples of the segmentation results of the methods to compare visually the advantages and disadvantages of them against each other. Although such a comparison is not comprehensive since the vessel segmentation performance in each image, depending on the conditions of that image, might be greater in one method or another, it can help to grasp how they work qualitatively.

\begin{figure}[ht!]
\begin{center}
\includegraphics[scale=0.5]{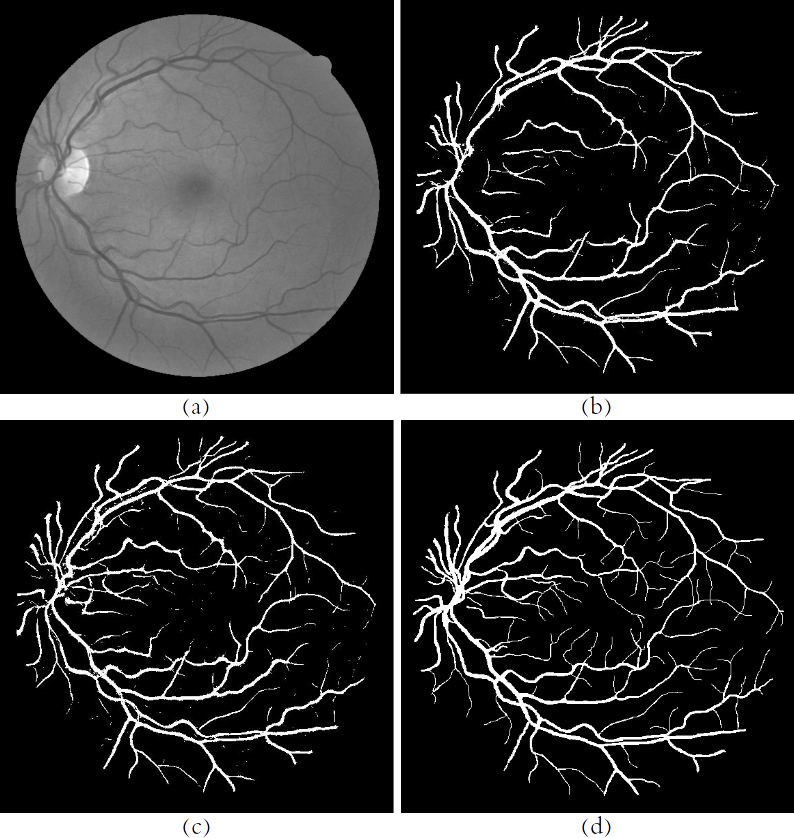}
\end{center}
\caption{An example of vessel segmentation in healthy images of DRIVE dataset. (a) The input image. (b) Vessel segmentation with Method 1, (TP-rate, FP-rate)=(0.77, 0.03), (c) Vessel segmentation with Method 2, (TP-rate, FP-rate)=(0.73, 0.02), (d) The corresponding ground truth image.}
%Source:
\label{5.5-image-DRIVE-healthy-Seg-Resutl}
\end{figure}

An example of the operation of the methods on a healthy image is shown in Fig. \ref{5.5-image-DRIVE-healthy-Seg-Resutl}. As can be seen, part (c) of this figure (Segmentation by Method 2) reveals more details of the vessels, which are not detected in Fig. \ref{5.5-image-DRIVE-healthy-Seg-Resutl} (b) (Segmentation by Method 1). Given the way the two methods work, such a difference is expected. Because the clustering process used in Method 1 to reduce bright lesions has no mechanism for maintaining linear structures and its sole purpose is to eliminate bright lesions only based on the anomalies of the illumination intensity.

On the other hand, Method 2 does a balance between eliminating bright lesions and not smoothing out the structures. Such a perspective in the two methods even is clear in identifying vessels within the optic disk. As can be seen, in the response of Method 2, the vessels within the optical disk have become more apparent. Of course, the larger sensitivity in maintaining linear structures of Method 2 has also led to the False-Positive pixels being added to the final image, an example of which can be seen in the over-detection of edges of the optic disc. In identifying the small vessels, the two methods have occasionally worked better than one another, but again in this regard, Method 2 is better in signifying details.

Furthermore, it can be seen that at locations where both methods have identified vessels, the vessels extracted in Method 1 are wider than vessels resulted in Method 2. The reason is that in the regularization process of Method 2, we use a kind of edge detection based on derivation, and where the changes are very gentle (vessel edges), this method does not produce a high response. Therefore, it is also difficult for the line operator to identify these edges. Obtaining the vessel pixels at the boundaries in Method 1 depends only on the line operator and the earlier stage does not do anything about it.

As is clear, both methods have worked well in categorizing Fovea and Macula as the background.

The image examined was one of the images of the DRIVE dataset, which was of optically good quality with no abnormalities. In the next part, we evaluate the segmentation procedure on one of the STARE images with bright lesions.

\begin{figure}[ht!]
\begin{center}
\includegraphics[scale=0.5]{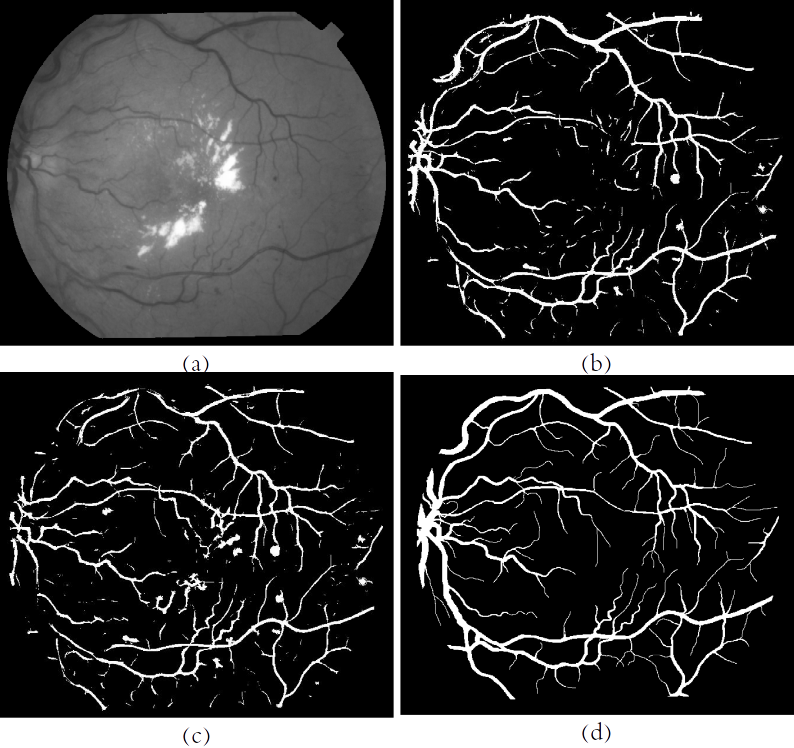}
\end{center}
\caption{An example of vessel segmentation in pathological image of STARE dataset. (a) The input image. (b) Vessel segmentation with Method 1, (TP-rate, FP-rate)=(0.73, 0.02), (c) Vessel segmentation with Method 2, (TP-rate, FP-rate)=(0.62, 0.03), (d) The corresponding ground truth image.}
%Source:
\label{5.6-image-STARE-pathological-Seg-Resutl}
\end{figure}

In Fig. \ref{5.6-image-STARE-pathological-Seg-Resutl}, the result of vessel segmentation by both methods is shown on one of the STARE images, which has both bright lesions and slightly optically heterogeneous edges. Obviously, in places where image light is not good (such as the top left corner of the image), Method 1 works better. Also, Method 1 has worked much better in this image to eliminate bright lesions. But elsewhere, as in the previous image, Method 2 has revealed more details.

\begin{figure}[ht!]
\begin{center}
\includegraphics[scale=0.5]{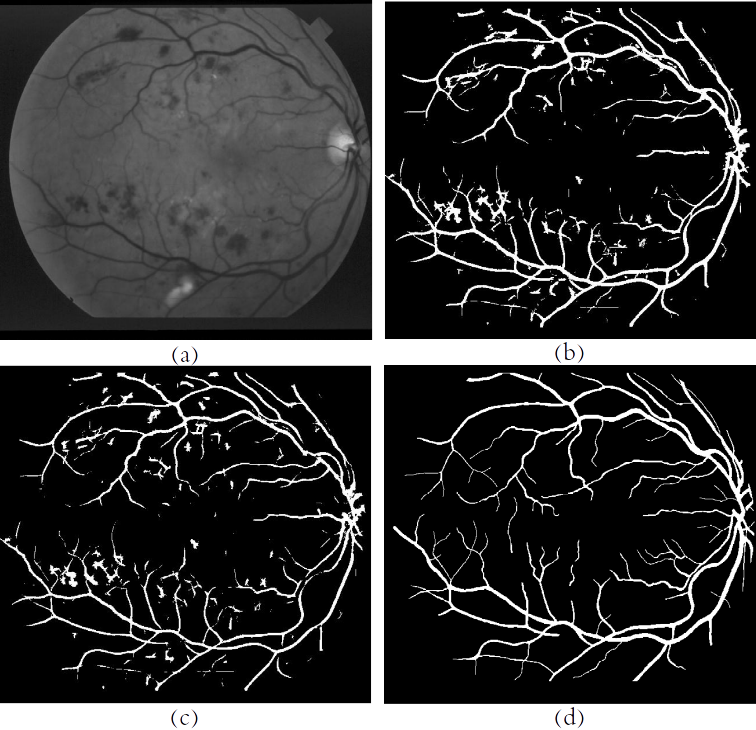}
\end{center}
\caption{An example of vessel segmentation in images with dark anomalies. (a) The input image. (b) Vessel segmentation with Method 1, (TP-rate, FP-rate)=(0.76, 0.04), (c) Vessel segmentation with Method 2, (TP-rate, FP-rate)=(0.86, 0.11), (d) The corresponding ground truth image.}
%Source:
\label{5.7-image-Dark-Anomaly-Seg-Resutl}
\end{figure}

Fig. \ref{5.7-image-Dark-Anomaly-Seg-Resutl}  is an example of a segmentation performed in an image with dark anomalies. 
As discussed in Chapter 4, although dark anomalies have the same light intensity as blood vessels, they are round and small in shape and different from the linear structure of vessels. They are also distributed irregularly across the retina image. Therefore, the line operator detector, which follows linear structures on the retina, can partly ignore these anomalies.

The performance of eliminating anomalies in this image for both methods is almost the same, since the anomalies of this image are reduced only in the line operation which were common between the two methods It is also clear that Method 2 has better identified the vessels within the optic disc.

\begin{figure}[ht!]
\begin{center}
\includegraphics[scale=0.5]{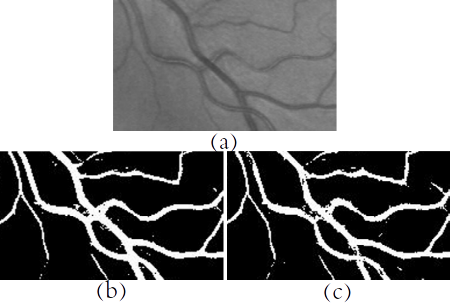}
\end{center}
\caption{An example of vessel segmentation in vessels with central reflection. (a) A portion of a DRIVE image with central reflection. (b) Vessel segmentation with Method 1, (c) Vessel segmentation with Method 2.}
%Source:
\label{5.8-image-Central-Reflx-Seg-Resutl}
\end{figure}

An example of a vessel segmentation is shown in Fig. \ref{5.8-image-Central-Reflx-Seg-Resutl}, which shows only a portion of the input image having a central reflection in the vessel, as well as the results of vessel segmentation for both methods in such areas. As can be seen, both methods have worked well in identifying vessel pixels in the centre. 

The best and worst segmentation results are presented in Fig. 5.9 and Fig. 5.10, respectively, by Method 1, and in Fig. 5.11 and Fig. 5.12 by Method 2, respectively, on the DRIVE and STARE datasets.

\begin{figure}[ht!]
\begin{center}
\includegraphics[scale=0.5]{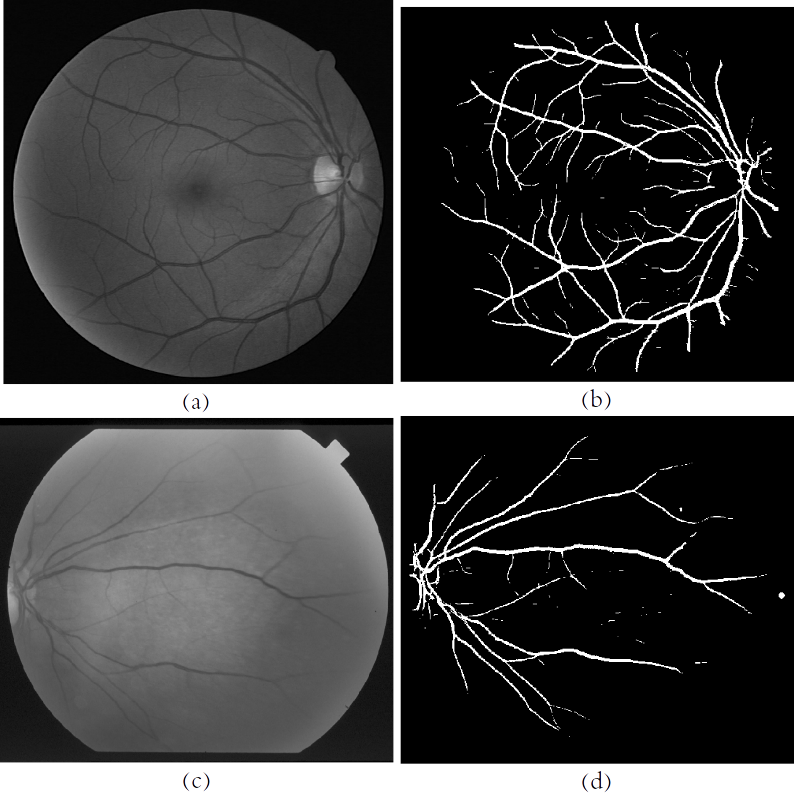}
\end{center}
\caption{The best results of Method 1. (a),(b) The DRIVE image and the corresponding segmentation result. (c),(d) The STARE image and the corresponding segmentation result.}
%Source:
\label{5.9-image-Best-Method1}
\end{figure}

\begin{figure}[ht!]
\begin{center}
\includegraphics[scale=0.5]{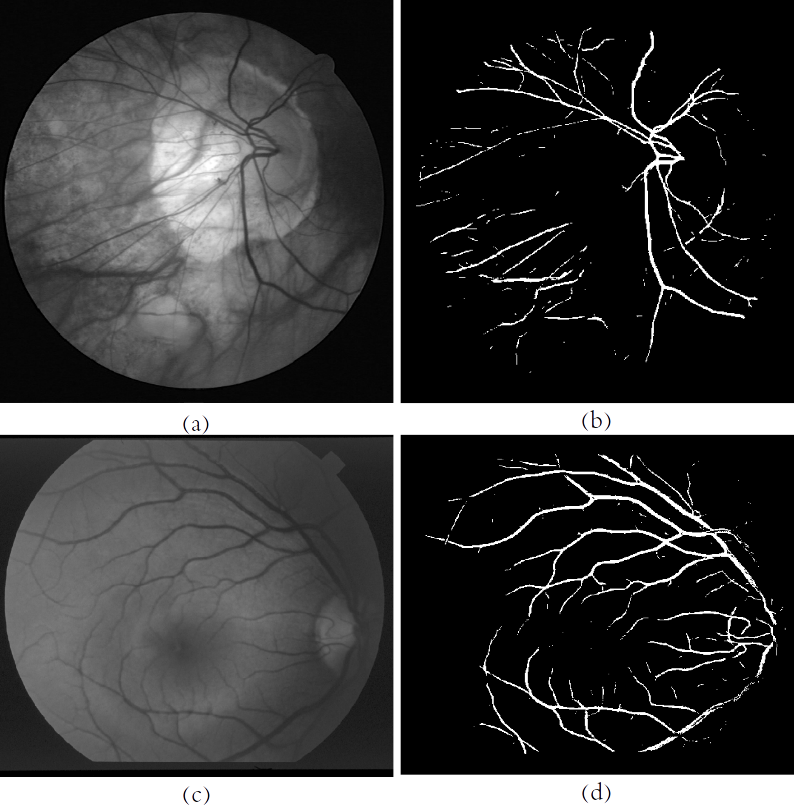}
\end{center}
\caption{The worst results of Method 1. (a),(b) The DRIVE image and the corresponding segmentation result. (c),(d) The STARE image and the corresponding segmentation result.}
%Source:
\label{5.10-image-Worstt-Method1}
\end{figure}

\begin{figure}[ht!]
\begin{center}
\includegraphics[scale=0.5]{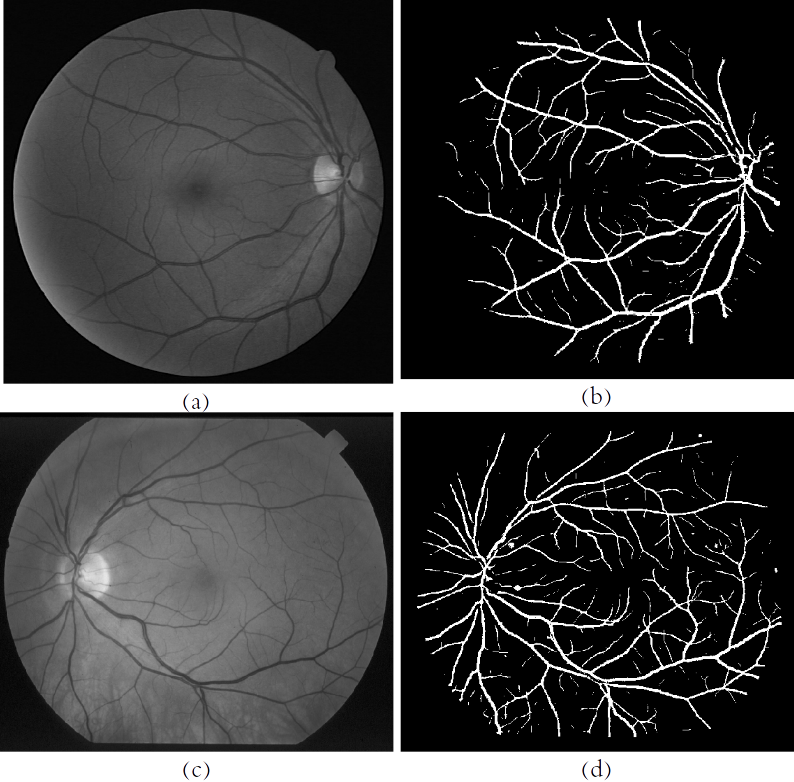}
\end{center}
\caption{The best results of Method 2. (a),(b) The DRIVE image and the corresponding segmentation result. (c),(d) The STARE image and the corresponding segmentation result.}
%Source:
\label{5.11-image-Best-Method2}
\end{figure}

\begin{figure}[ht!]
\begin{center}
\includegraphics[scale=0.5]{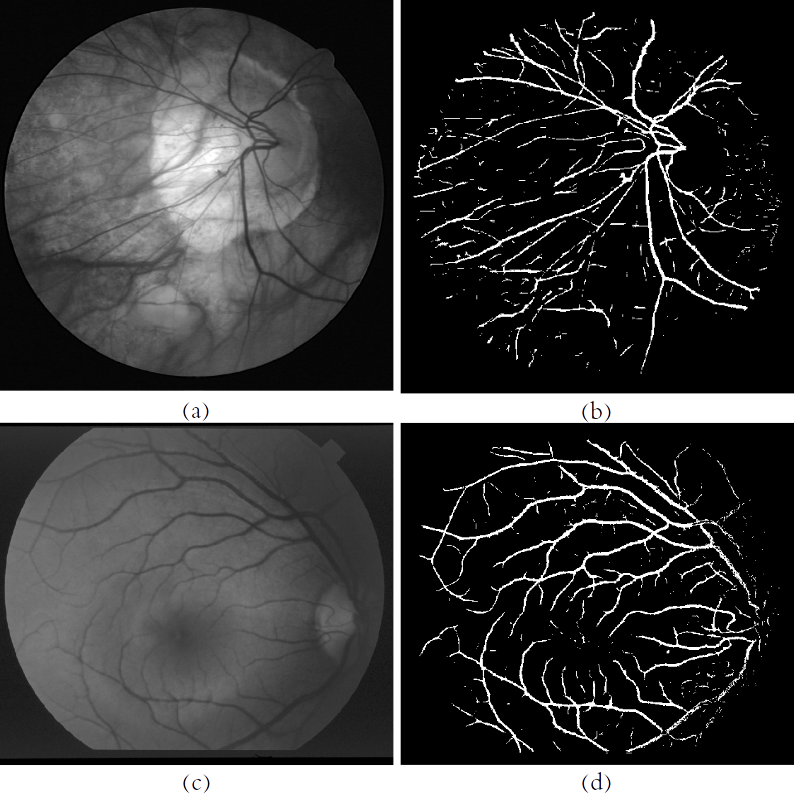}
\end{center}
\caption{The worst results of Method 2. (a),(b) The DRIVE image and the corresponding segmentation result. (c),(d) The STARE image and the corresponding segmentation result.}
%Source:
\label{5.12-image-Worst-Method2}
\end{figure}

Only from the qualitative analysis, ignoring some of the vessel pixels in Method 1, especially in the optic disk, and categorizing some non-vessel pixels as vessel pixels in Method 2, can be considered as deficiencies of the two methods in comparison with each other. Therefore, one cannot be completely superior to the other. This similarity in performance was also seen in the results of the statistical evaluations.

This chapter dealt with the statistical evaluation of the methods using the accuracy, area under the ROC curve and response time.

In addition to the statistical perspective, the methods were compared from a qualitative perspective and the characteristics of their vessel segmentation responses were investigated. 
Based on what has come out in this chapter, the methods in this thesis in the field of retinal vessel segmentation have acceptable performance in comparison with other methods.

\chapter{Conclusion and Future Works}

\section{Conclusion}
In this thesis, two automated methods for detecting vessels in retinal images were presented, with the aim of providing decision support systems for physicians. The features of the vessels and anomalies, as well as the challenges facing the segmentation, were investigated.

The basis of both approaches is one view, that of eliminating bright lesions, prior to the final vessel segmentation procedure. The final segmentation is performed by a multi-scale line operator. This operator detects vessels based on their most obvious feature, namely linearity.

The efficiency of the two methods in identifying the vessels, based on the accuracy and area under the ROC curve, was obtained by comparing them with the labelled images. Examination of all three images of the DRIVE and STARE datasets showed that both methods have good performances and response times in detecting vessels in healthy and pathological images.

All together suggested methods have the following features:

\begin{itemize}
\item Detection of vessels based on their structural features and retinal image conditions;
\item Performances comparable to other methods;
\item Eliminating the need to set parameters for images with different conditions thus increasing generalization;
\item Effective response time in reducing abnormalities and detecting vessels simultaneously.
\end{itemize}

\section{Suggestions for Future Works}

In any area of research, there are always ways to improve that have not yet been explored. Finding such solutions can lead to greater efficiency in any field. Following are the methods suggested in this thesis:
\begin{itemize}
\item Using better and more task-specific regularization methods to find vessels and reduce dark and bright anomalies;
\item Applying vessel detection techniques along with other methods for identifying other parts of the retina, such as optic discs, to produce a therapeutic system;
\item Parallelizing procedures of the algorithms to reduce response time and achieve a real-time medical identification system;
\item Using other methods to identify linear and curved structures.
\end{itemize}
The described algorithms, although tested only on retina images, can also be reproduced for other medical and non-medical applications.

% Appendices
\appendix

%%%%%%%%%% DON'T DELETE THIS, REVERTS NUMBERING BACK %%%%%%%%%%%%%
\makeatletter
\renewcommand{\@makechapterhead}[1]{\vspace *{-10\p@ }{\parindent \z@ 
\raggedright \normalfont \ifnum \c@secnumdepth >\m@ne \Huge \bfseries 
\@chapapp \space \thechapter \vskip 10\p@ \fi #1\par \nobreak \vskip 30\p@ }}
\makeatother
%%%%%%%%%% DON'T DELETE THIS, REVERTS NUMBERING BACK %%%%%%%%%%%%%

\chapter{Weber's Law and Space}
\section{Weber's Law}
As a result of the work by E. H. Weber and Jay. T. Fechner the Weber-Fechner Law, or the Weber Law introduced. This law represents the universal influence of sensory stimulus on the background of the environment on human sensitivity of changes, or Just-Noticeable-Difference (JNU) related to an optical and acoustic stimulus. Weber's equation is shown to be equal a constant:

\begin{equation}
    \frac{\delta u}{u} = Weber{ }constant
\end{equation}

For a qualitative perspective of Weber's law, examples of it can be examined in everyday events. In a crowded stadium where the background noise is high, if one wants to be heard (to be recognizable), she has to speak loud enough that it sounds like a shout. The same thing can be seen in other ways in visual phenomena. In a dark night in the absence of a bright moon and neon lights, one can easily see the stars in the sky. Otherwise, it would be difficult for the stars to be seen.

In this dissertation, we used Weber's Law of vision to increase the efficiency of our system \cite{shen2002weber}.

\section{Weber's Law in Image Processing}
The relation (A.1) can also be written as follows:
\begin{equation}
    \frac{\delta f(z)}{f(z)} = k^*
\end{equation}

In which \(f(z)\) is the magnitude of the physical illumination, \(\delta f(z)\) is the rate of change in the brightness and \(k^*\) is the Weber constant.

If in the pixel \(z=z^*\) the value of the left size is less than \(k^*\), human's eye cannot detect the change in the magnitude of the physical intensity. This means that the intensity variations of the image intensity are much less than the background intensity. Depending on the intensity of the perceived illumination, \(v^0\), the rate of change of its intensity is related to the magnitude of the physical illumination intensity:

\begin{equation}
    \frac{\delta f(z)}{f(z)} = k \times \delta v^0
\end{equation}

Obviously, if \(\delta v^0\) is small, it is difficult to detect the change in physical illumination intensity, and vice versa. In image processing, the magnitude of the physical intensity is the intensity of a two-dimensional image. In these processes, the rate of change, \(\delta f(z)\) is known as the intensity gradient. The transformation of the input image,\(I^0\) to the intensity of the received illumination's space (Weber Space) can be obtained by the following theory.

\textbf{Theory}: We have, \(v^0(0,0) = 0\) and \(f(0,0) = 1\). Having the function of the physical intensity magnitude, \(f(x,y)\), in (A.3), when the rate of change, \(\delta f(x,y)\), is small for the pixel at \((x,y)\), the perceived intensity, \(v^0(x,y)\), can be obtained as follows (Proof in \cite{lam2010general}):
\[v^0(x,y)=\frac{ln(f(x,y))}{k}\]In the above theory, two assumptions are made; \(v^0(0,0)=0\) and \(f(0,0)=1\). If \(f(z)=1+I^0(z)\) and \(z=(x,y)\), these assumptions would be held true. In a retina image, the image boundary pixels (outside the field of view) have zero intensity, ie, \(I^0(0)=0\) and we have \(f(0)=1\) and \(v^0(0)=ln(1+0)=0\). So, the formula that transforms the input image,\(I^0\), to the Weber Space image, \(v^0\) will look like this:
\begin{equation}
	v^0(z) = \frac{ln(1+I^0(z))}{k}
\end{equation}The constant parameter,\(k\), is taken 1 here \cite{lam2010general}.

\chapter{Total Variation Model and Digital Total Variation Filter}

\section{Total Variation Model}

As mentioned earlier, the Total Variation (TV) model is used to eliminate noise in signals (such as images). We will further explore this model.

Assume that \(u^0(x,y)\), \(x,y \in \Omega\) is a noisy image and \(u(x,y)\) is a desirable and unnoisy image. Then we have:
\begin{equation}
	u^0(x,y) = u(x,y)+n(x,y)
\end{equation}Where \(n\) is the noise with zero mean and variance \(\sigma^2\) \cite{chan2001digital}. 
\[En(x)=0, En^2(x)=\sigma^2\]
Our goal is to reconstruct \(u(x,y)\) from \(u^0(x,y)\) \cite{rudin1992nonlinear}. The Constrained Minimization (Optimization) Problem \cite{rudin1994total} is concerned with the changes in an image. These changes reflect the extent to which the image under study is oscillatory. This optimization problem can be described as follows:
"Minimize the following relation subject to the constraints of \(u^0\)." \[
\int_\Omega \sqrt{(u_x^2+u_y^2)}
\]
This relation is the sum of the variations in the intensities of an image and can be illustrated as follows \cite{chan2001digital}:
\begin{equation}
	TV[u]=\int_\Omega |\nabla u|dxdy
\end{equation}Where \(\Omega\) is the space of the pixels of the image and \(|\nabla u|\) is the gradient's value.
One of the stated constraint is on the mean:
\begin{equation}
\int_\Omega udxdy = \int_\Omega u^0dxdy
\end{equation}This limitation is obtained from the zero mean noise, \(n(x,y)\) in (B.1).
Another constraint is on the noise variance:
\begin{equation}
\frac{1}{|\Omega|}\int_\Omega (u-u^0)^2dxdy=\sigma^2
\end{equation}Where, \(\sigma^2\), the noise variance, is estimated from the image.
Clearly, minimizing the sum of changes results in a reduction in the image and consequently a reduction in noise. But we know that this change in addition to noise cancellation also brings other details down. Hence, there is a need for a fit factor alongside the model of sum change. Conditions (B.3) and (B.4), as well as the Total Variation Model in relation (B.2) together, provide a constrained optimization problem.
Since \(TV[u+c]=TV[u]\) is true for any fixed number, \(c\), the first constraint holds (B.3). So we have to consider only the second constraint. The following energy function can be defined by introducing the Lagrange coefficient \(\lambda\):
\begin{equation}
	J[u]=\int_\Omega |\nabla u|dxdy+\lambda \int_\Omega (u-u^0)^2dxdy
\end{equation}
The Euler-Lagrange equation of this is:
\[\nabla . \frac{\nabla u}{|\nabla u|}+\lambda (u^0-u)=0\]By solving this equation using TV filter, the optimum signal (image), \(u\) is obtained.

The above equations are studied in continuous space. What follows is the function of the filter of the sum of changes in discrete space, and the only difference in relationships is the use of the sigma symbol instead of the integral symbol.

\section{Total Variation Filter}
Filter The sum of discrete changes is a filter for filtering and improving images or data that can be represented by a graph. This filter is bypass and can filter images without blurry edges. Applying this filter in an iterative process solves a Global Total Variation (\(L^1\)) problem. Its applications are given for the noise of 1D signals, 2D data with irregular structure and grey and colour images.
Features of this filter can be listed as follows:
\begin{enumerate}
\item This filter has a fixed, simple structure for the coefficients that contain information about the edges themselves (so this filter is data-dependent).
\item Mathematically, unlike most statistical filters, the sum of the changes is based on geometric and functional analysis.
\item From a practical point of view, this filter is very consistent in working with graphical data and vector signals, such as images.
\end{enumerate}
The motivation for using the filter is the change in image processing, its ability to reduce noise
And more useful is the ability of the filter to highlight homogeneous segments including edges \cite{sawatzky2011nonlocal}.

In the following sections, we will explain how this filter works and how the formulas are used.
\section{Graphs and the Total Variation Filter}
Here, the filter structure is called the sum of discrete changes and their properties. Noise data, which must be recorded, is on a graph. First, consider some differential formulas on graphs.
\subsection{Graphs and Edge Derivatives} 
A discrete space can be modelled with a graph \([\Omega,E]\), which contains a set of nodes,\(\Omega\), and a set of edges, \(E\). The scalar signal \(u\) is a function on \(\Omega\):
\begin{equation}
	u:\Omega \to R
\end{equation}The value of pixel \(\alpha\) value is given by \(u_\alpha\). The local variation or strength in each pixel \( \alpha\) is defined as follows:
\begin{equation}
|\nabla u|:= \sqrt{\sum_{\beta {\sim} \alpha}(u_\beta - u_\alpha)^2 }
\end{equation}For any number \(a\) greater than zero, the Regularized Local Variation is as follows:
\begin{equation}
	|\nabla_\alpha u |_a = \sqrt{|\nabla_\alpha u|^2 + a^2}
\end{equation}Where \(a\) is a very small number.
After that, the Edge Derivative is defined. \(e\) represents the edge \(\alpha \sim \beta\). The edge derivative of \(u\) in the direction of \(e\) and in the pixel \(\alpha\) is:
\begin{equation}
\frac{\partial u}{\partial e}:=u_\beta - u_\alpha
\end{equation}And it will be computed:
\begin{equation}
	|\nabla_\alpha u | = \sqrt{\sum_{e - \alpha}[\frac{\partial u}{\partial e}|_\alpha]^2}  ,
	 \frac{\partial u}{\partial e}|_\alpha:=-\frac{\partial u}{\partial e}|_\beta
\end{equation}Where \(e-\alpha\) means that \(\alpha\) is of one of the nodes of the edge \(e\).

Two graphical structures for use in image processing are shown in Fig. B.1.

\begin{figure}[ht!]
\begin{center}
\includegraphics[scale=0.5]{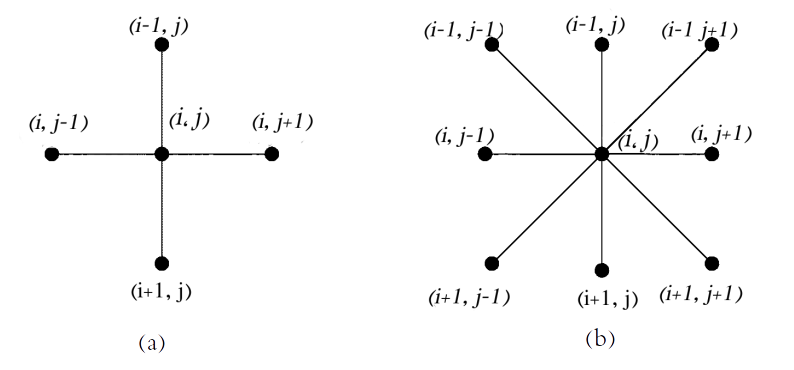}
\end{center}
\caption{Two graph structures for discrete and square domain of images (the indexing is based on MATLAB numbering system.) In (a) a five-node pattern is shown in which each pixel \(\alpha=(i,j)\) has four neighbours, but in (b) it has eight neighbours \cite{chan2001digital}}
%Source:
\label{B.1-image.TV-Graph}
\end{figure}

\subsection{Digital Total Variation (DTV) Filter}
The DTV has two parameters:
\begin{itemize}
\item A small number \(a\) greater zero called Regularization Parameter (B.8);
\item A parameter \(\lambda\) greater than zero, called the Fitting Parameter or Lagrange Coefficient (determines the effect of the similarity of the two images in the regularization).
\end{itemize}
The parameter \(a\) is to prevent zero denominator in formulas B.7 and B.8. If it is small, the TV filter efficiency does not depend on it. However, the parameter \(\lambda\) is important for the regularization between denoising and smoothing and it depends on the amount of the noise in the image (how to set this parameter is explained in Section B.5).
For a noisy signal, \(u^0\),, the TV filter, \(F^{\lambda,a}\), is a nonlinear and data-dependent filter.
\[F^{\lambda,a}:u \to v\]Here, \(u\) is any signal on the space \(\Omega\) and \(v\) is the filter's output. For simplicity, \(F^{\lambda,a}\) will be replaced by \(F\). For each node \(\alpha\) in the space \(\Omega\) (Fig. B.2) we have:
\begin{equation}
v_\alpha = F_\alpha(u)=\sum_{\beta \sim \alpha}h_{\alpha \beta}(u)u_\beta+h_{\alpha \alpha}(u)u_\alpha^0
\end{equation}Here the coefficients are:
\begin{equation}
h_{\alpha\beta}=\frac{\omega_{\alpha\beta}(u)}{\lambda+\sum_{\gamma \sim \alpha}\omega_{\alpha\beta}(u)}\  ,
h_{\alpha\alpha}=\frac{\lambda}{\lambda+\sum_{\gamma \sim \alpha}\omega_{\alpha\beta}(u)}\  ,
\omega_{\alpha\beta}=\frac{1}{|\nabla_\alpha u|_a}+\frac{1}{|\nabla_\beta u|_a}
\end{equation}
The TV filter algorithm will be as follows:
Perform the filter in one pixel:
\begin{enumerate}
\item The Digital TV filter in pixel \(\alpha\):
\begin{itemize}
\item Obtain the change (derivative) in the neighbourhood, \(|\nabla u|_a\), in Pixel \(\alpha\) and in all its neighbours, \(\beta\),
\item Compute the weights \(\omega_{\alpha\beta}\) and filter's coefficients, \(h_{\alpha\beta}\) and \(h_{\alpha\alpha}\), by formulas B.12.
\end{itemize}
\end{enumerate}
Perform the filter to the whole image:
Put the pixels in a sequence;

\begin{enumerate}
\item The TV filtering process on the whole image:
\begin{itemize}
\item Put \(u^{(0)} = u^0\) (initialization);
\item Compute \(u\) in all nodes \(\alpha\): \[u^{(k)}_{\alpha_j} = F_{\alpha_j}=(u^{(k-1)})\]
\item Repeat the previous step until the process converges.
\end{itemize}
\end{enumerate}

If the digital TV filter converges, the finite signal, \(u\) will be the only minimizer for the TV energy (a discrete form of the relation (B.5)):

\[FTV[v]=\sum_{\alpha \in \Omega}|\nabla_\alpha v|_a+\lambda\sum_{\alpha \in \Omega}(v_\alpha-v_\alpha^0)^2\]\section{Lagrange coefficient}
This coefficient is used to establish the constraint (B.4) in the Euler-Lagrange relation (B.5). The \(\lambda\) parameter plays an important role in the regularization between smoothing and the difference between the output image and its original quality. This parameter determines the effect of the similarity difference between the two images in the regularization. There are some ways to estimate it. From the discrete model perspective, an optimal estimate of \(\lambda\) of the current signal \(u\) is:
\[\lambda=\frac{1}{\sigma^2}\frac{1}{|\Omega|}\sum_{\alpha \in \Omega}\sum_{\beta \sim \alpha}\omega_{\alpha\beta}(u_\beta-u_\alpha)(u_\alpha-u_\alpha^0)\]Here, \(\sigma^2\), is the noise variance, which can be estimated from the image. \(\omega_{\alpha\beta}\) are the filter's weights introduced in (B.12). \(|\Omega|\)is the node space.

This parameter is updated for the image every few iterations (for example every 2 iterations) of the TV filtering. In some applications, they have used a fixed size for this parameter. Further explanation and proof of the formulas can be found in \cite{chan2001digital}.

% Bibliography
% Bibliography
\begingroup
    \setlength\bibitemsep{10pt}
    \linespread{1}\selectfont
    \printbibliography[title=REFERENCES]
\endgroup
\addcontentsline{toc}{part}{REFERENCES}

\end{document}